\pdfoutput=1
\documentclass[twocolumn,10pt,a4paper,floatfix,prl,unsortedaddress,superscriptaddress,longbibliography]{revtex4-1}

\usepackage[utf8]{inputenc}
\usepackage{microtype}
\usepackage{newtxtext}
\usepackage[upint]{newtxmath}
\usepackage{eucal}

\usepackage{graphicx}
\usepackage{booktabs}
\usepackage{enumerate}
\usepackage{siunitx}
\usepackage{braket}
\usepackage{color}
\usepackage{float}
\usepackage{soul}
\usepackage{bm}
\usepackage{tikz}
\usetikzlibrary{decorations.pathreplacing}

\usepackage[colorlinks,allcolors=blue]{hyperref}
\usepackage[capitalize]{cleveref}

\newcommand{\vecq}{{q}}
\newcommand{\veck}{{k}}
\newcommand{\vecQ}{{Q}}

\newcommand{\vecS}{\boldsymbol{S}}

\newcommand{\vecdQ}{\Delta{Q}^{\uparrow \downarrow}}
\newcommand{\vectau}{{\tau}}

\definecolor{DarkRed}{rgb}{0.80,0,0}
\definecolor{DarkBlue}{rgb}{0.20,0,0.80}
\definecolor{Purple}{rgb}{0.55,0,0.55}
\definecolor{Magenta}{rgb}{1,0,1}

\begin{document}
\title{Designing lattice spin models and magnon gaps with supercurrents}
\author{Johanne Bratland Tjernshaugen}
\affiliation{Center for Quantum Spintronics, Department of Physics, Norwegian \\ University of Science and Technology, NO-7491 Trondheim, Norway}
\author{Martin Tang Bruland}
\affiliation{Center for Quantum Spintronics, Department of Physics, Norwegian \\ University of Science and Technology, NO-7491 Trondheim, Norway}
\author{Jacob Linder}
\affiliation{Center for Quantum Spintronics, Department of Physics, Norwegian \\ University of Science and Technology, NO-7491 Trondheim, Norway}

\begin{abstract}
\noindent Electric control over magnetic interactions at the level of individual spins is relevant for a variety of quantum applications, such as qubits, memory and sensor functionality. We show here that spin lattices and magnon gaps can be controlled with a supercurrent. Remarkably, a spin-polarized supercurrent makes the interaction between magnetic adatoms placed on the surface of a superconductor depend not only on their relative distance, but also on their absolute position in space. This property permits electric control over the interaction not only between two individual spins, but over an entire spin lattice, allowing for tunable non-collinear ground states and a practical arena to study the properties of different spin Hamiltonians. Moreover, we show that a supercurrent controls the magnon gap in antiferromagnetic and altermagnetic insulators. These results provide an accessible way to realize electrically controlled spin switching and magnon gaps without dissipative currents. 
\end{abstract}

\maketitle

\textit{Introduction}--- Electrical control of magnetic interactions is relevant for a variety of quantum applications, such as sensing, computation and memory functionality. Quantum sensing with spin defects \cite{roberts2025quantum, fang2024quantum} allows for optically measuring nuclear magnetic resonance (NMR) \cite{fang2024quantum}. Quantum computing requires interactions between qubits, such as the magnetic Ruderman–Kittel–Kasuya–Yosida (RKKY) interaction connecting spin qubits \cite{tanamoto2021compact, yang2016long, mishra2021yu}. Finally, memory functionality based on electric writing of the spin-state improves the efficiency of such devices \cite{zhang_sciadv_24}.

The RKKY interaction mediated by electrons in a superconductor provides one promising way of controlling the coupling between spins at an individual level \cite{nadj2014observation, pupim2025adatom, kim2018toward, heimes2015interplay, mohanta2018supercurrent, yazdani1997probing, rontynen2015topological}. Of particular interest is typically the degree of spin non-collinearity, as this gives rise to interesting spin textures such as spin spirals \cite{kimura2007spiral, tokura2010multiferroics}, skyrmions \cite{nagaosa2013topological, zhang2020skyrmion, marrows2021perspective} and non-collinear antiferromagnetism \cite{rimmler2025non-collinear}, all with potential spintronic applications. The ground state configuration of an ensemble of spins can be non-collinear due to a Dzyaloshinskii-Moriya (DM) interaction induced by spin orbit coupling \cite{soumyanarayanan2016emergent} or from mixed-parity superconductivity with $s+ip_x$ symmetry \cite{ouassou2025dzyaloshinskii}.  Another quantity of interest in an ensemble of interacting spins is the magnon energy gap \cite{chumak2015magnon, oba2015electric, pradipto2017mechanism, yoshii2020detection, zhu2021topological, yumnam2024magnon, mardele2024tuning}. A tunable magnon gap can, as an example, give rise to a tunable lattice thermal conductivity \cite{vu2023magnon}. The interaction of magnons with superconducting qubits \cite{tabuchi2015coherent, tabuchi2016quantum} and flux quanta in superconductors \cite{dobrovolskiy2019magnon, niedzielski2023magnon} makes the generation and tuning of magnons possible in superconducting hybrid systems.  

\begin{figure*}
    \centering
    \includegraphics{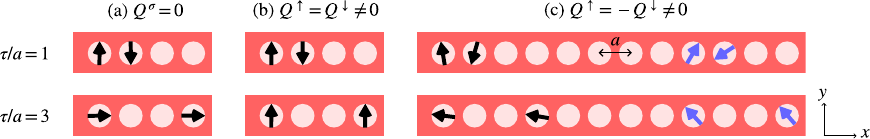}
    \caption{Two spins $\vecS_1$ and $\vecS_2$, here represented as large arrows in the $xy$--plane, are placed on the top of a 1D superconductor with equal-spin Cooper pairs. The circles represent the lattice points, and the lattice constant is $a$. The interaction between the two spins, and hence their ground state configuration, can be tuned by a supercurrent. The positions of the spins are $r_1,r_2$ and the relative coordinate is $\tau = r_2-r_1$. (a) shows the ground state configuration for two spins separated by a distance $\tau/a=1$ and $\tau/a=3$ when the supercurrent is zero. When a charge current $Q^{\sigma}/Q_c\approx0.87$ 
    is turned on in (b), the ground state changes for $\tau/a=3$.
    In (c), a spin current $Q^{\sigma}/Q_c\approx0.87\sigma$
    flows through the superconductor. This alters the ground state compared to (a) and (b), and the spins are allowed to be non-collinear.
    Moreover, the spin current creates a dependence on the center-of-mass coordinate of the spins, unlike conventional spin-spin interactions that only depend on the relative coordinate. The blue arrows represent the ground state of the spins when they are moved 8 lattice sites to the right. 
   The chemical potential is set to $\mu=-t$. The supercurrent is normalized on the critical supercurrent $aQ_c=0.0505$, which is found by solving the gap equation. }
    \label{fig:spinlattice}
\end{figure*}

In this Letter, we show that supercurrents in superconductors enable electric control over both magnon gaps and magnetic interactions in a way that permits controllable design of a spin lattice. We use perturbation theory to show analytically that a spin supercurrent carried by triplet Cooper pairs gives rise to a number of different interaction terms between magnetic adatoms placed on the superconductor \cite{liebhaber2022quantum}. This causes the ground state configuration of the spins to be non-collinear. Importantly, we find that the ground state depends on the center-of-mass coordinate of the spins, unlike the usual RKKY interaction. This property allows control over the interaction between two spins as illustrated in Fig. \ref{fig:spinlattice}, and in fact over an entire spin lattice, providing an arena to study the properties of different spin Hamiltonians.
Moreover, we demonstrate that when an antiferromagnetic or altermagnetic insulator is placed on top of the superconductor, the supercurrent controls the magnitude of the magnon gap.
This can be thought of as a dissipationless magnon transistor and is achieved even with a conventional singlet BCS superconductor. The magnon gap correction is a first-order effect in the coupling between spins and quasiparticles in the superconductor, while the spin-spin interactions enabling spin lattice design appear at second order. The electrical control of a spin lattice and the magnon gap via supercurrents predicted in this Letter reveals a synergy between superconductivity and magnetism that opens new research avenues to explore. 

\textit{Effective spin-spin interaction}---%
 For simplicity, we consider an effective one-dimensional (1D) superconducting chain with magnetic adatoms placed on its surface, as this will permit us to numerically simulate large systems and since experiments often can be modeled well by effective 1D Hamiltonians \cite{liebhaber2022quantum, lesueur_prl_08}. Our results carry over straightforwardly to 2D and 3D systems \cite{SM}.  
The full Hamiltonian for the system is $H= H_\text{SC}+H_c$, where {$H_\text{SC}$ is the Hamiltonian for the superconductor and $H_c$ is the coupling between the magnetic adatoms and the electrons in the superconductor. 
We consider a superconductor with equal-spin triplet Cooper pairs: this scenario applies both to induced triplet superconductivity via the proximity effect and to an intrinsic triplet superconductor. The former can be achieved by proximitizing an $s$-wave superconductor with one or more ferromagnets \cite{halterman2007odd, halterman2008induced, kalcheim_prb_15, grein2013inverse, eschrig2008triplet, khaire_prl_10, robinson_science_10} or a conductor with Rashba spin-orbit coupling \cite{reeg2015proximity}, curving the superconductor \cite{ying2017tuning}, or by irradiating 
it with light \cite{gassner2024light}.
The starting point is the attractive extended Hubbard model in real space,
\begin{align}
    H_\text{SC} =-\mu\sum_{i,\sigma}n_{i,\sigma} - \sum_{\langle i,j\rangle,\sigma} \Big( 2t c_{i,\sigma}^{\dagger} c_{j,\sigma}
    +\frac{U}{4}n_{i,\sigma} n_{j,\sigma}\Big)
\end{align}
where $n_{i,\sigma} = c_{i,\sigma}^\dag c_{i,\sigma}$, $\mu$ is the chemical potential, $t$ is the hopping constant between nearest neighbors, and $U>0$ is an attractive nearest-neighbor interaction. The operators $c_{i,\sigma}^{\dagger}$ and $c_{i,\sigma}$ creates and destroys, respectively, an electron with spin $\sigma$ at lattice site $i$. $\langle i,j \rangle$ denotes a sum over nearest-neighbor pairs without double counting. 
We apply the mean-field approximation to the last term in $H_\text{SC}$. The real-space order parameter (OP) is defined as $\Delta_{i,j}^{\sigma}=-U\langle c_{j,\sigma}c_{i,\sigma}\rangle/4$, and to model a supercurrent-carrying state we write it as \cite{Takashima2017:PRB}
 $\Delta_{i,j}^{\sigma}= \Delta_{\delta}^{\sigma} \text{e}^{\text{i}\vecQ^{\sigma}\cdot({r}_i+{r}_j)},$ where $\delta=r_j-r_i$ and $i,j$ are neighboring lattice sites. Adding a phase winding to the order parameter in this way describes conserved supercurrent flow in each spin channel. Thus, the magnitude $|\Delta_{\delta}^{\sigma}|$ does not depend on the positions $r_i$ and $r_j$ themselves. The phase gradient across the material allows for a supercurrent, and $Q^{\sigma}$ is the momentum of the spin--$\sigma$ Cooper pairs. If $Q^{\sigma} \neq Q^{-\sigma}$, the supercurrent carries charge and is spin polarized along the spin quantization axis $z$. If $Q^{\sigma} =- Q^{-\sigma}$, the system carries a pure spin supercurrent. 
We impose periodic boundary conditions, thus restricting the supercurrent momentum values to $Q^{\sigma} = 2\pi m/aN$ where $m$ is an integer and $N$ is the number of lattice points. 
Next, we Fourier transform the electron operators $c_{i,\sigma}^{(\dagger)}=N^{-1/2}\sum_k c_{k,\sigma}^{(\dagger)}\text{e}^{(-)\text{i}kr_i}$ and define the order parameter in $k$--space as $\Delta_k^{\sigma}=\sum_{\delta}\Delta_{\delta}^{\sigma} \text{e}^{\text{i}\veck\delta}$. This gives the Hamiltonian $H_\text{SC} = \sum_{\veck, \sigma} \epsilon_{\veck} c_{\veck, \sigma}^{\dagger} c_{\veck, \sigma}$ $+ \frac{1}{2}\sum_{\veck, \sigma} \left(\Delta_{\veck}^{\sigma}\right)^{\dagger}c_{-\veck+\vecQ^{\sigma},\sigma}c_{\veck+\vecQ^{\sigma}, \sigma} \text{+ h.c.}$
up to an irrelevant constant. All $k$-sums run over the crystallographic Brillouin zone. On a 1D lattice with lattice constant $a$, $\epsilon_{\veck} = -\mu-2t\cos(ak)$ and $\Delta_{\veck}^{\sigma}=\Delta^\sigma\sin(ak)$. 
The magnitude $\Delta^\sigma$ of the superconducting OP is calculated by solving the gap equation \cite{SM}, and we choose $U$ such that $\Delta_0^\sigma=0.1t$ at zero temperature.
The coupling term is generally given by 
\begin{equation}
   H_c=\frac{\Lambda}{{2}N}\sum_{\sigma, \sigma'}\sum_{\veck, \vecq} \sum_{i}\vecS_i\cdot\boldsymbol{\sigma}_{\sigma, \sigma'}\text{e}^{-\text{i}\vecq\cdot{r}_i} c^\dagger_{\veck+\vecq,\sigma}c_{\veck, \sigma'}.
\end{equation}
Here, $\Lambda$ is the $s-d$ coupling strength, and $\boldsymbol{\sigma}$ is the Pauli vector. Physically, the exchange $\Lambda$ can be modified by changing the overlap between the wavefunctions of the quasiparticles and the magnetic adatoms. To derive the effective spin-spin interaction, the magnetic adatoms are treated as classical spins $\vecS_i$ at site $i$ of the superconductor.

\begin{figure*}[t!]
    \centering
    \includegraphics{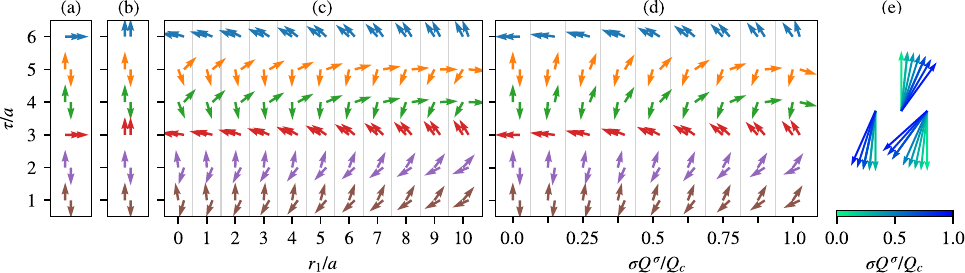}
    \caption{(a-d) The ground state configuration in the $xy$-plane of two spins $\vecS_1$ and $\vecS_2$ whose interaction is mediated by a superconductor with equal-spin Cooper pairs. Spins of the same color are separated by $\tau/a$. 
    (a) There are no supercurrents in the superconductor, $\vecQ^{\sigma}=0$. The ground states at $\tau/a=3$ and $\tau/a=6$ are degenerate in the $xz$-plane with negligible $y$-components (not shown here). (b) A charge current $Q^{\sigma}/Q_c\approx0.87$ flows through the superconductor. In both (a) and (b), there is no center-of-mass-dependence $(r_1)$. (c) The position of $\vecS_1$ is given by $r_1/a$, and the vertical gray lines separate spin pairs with different $r_1/a$. A spin supercurrent $Q^{\sigma}/Q_c\approx0.87\sigma$ flows through the superconductor and makes the RKKY interaction depend on the center-of-mass coordinate of the two spins. (d) Spin supercurrent dependence of the magnetic ground state for $r_1/a=8$. (e) Spin supercurrent dependence for a trimer with spin positions $r_1/a=4$, $r_2/a=6$ and $r_3/a=8.$ 
    The chemical potential is $\mu=-t$.
    }
    \label{fig:p-wave ground state}
\end{figure*}

We proceed to first diagonalize the mean-field Hamiltonian $H_\text{SC}$ and then perform a Schrieffer-Wolff transformation on $H$. We treat $H_c$ as the perturbation, and average out the fermions in the superconductor to get a pure spin Hamiltonian \cite{SM}. For a pure charge or spin supercurrent, there is no coupling between the magnetic adatoms and the superconductor to first order in $\Lambda.$ In this case, the effective spin Hamiltonian is of second order in $\Lambda$, and we find it to be
\begin{equation}\label{eq:effective spin-spin model p-wave}
    \begin{split}
        \mathcal{H}_\text{eff} = \sum_{i,j} \bigg(   J_{i,j} \vecS_i\cdot\vecS_j +   K_{i,j} S_i^z S_j^z +  D_{i,j}[\vecS_i\times \vecS_j]_z \\ +   L_{i,j}(S_i^x S_j^x - S_i^y S_j^y)+  M_{i,j}(S_i^x S_j^y + S_i^yS_j^x) \bigg).
    \end{split}
\end{equation}
A detailed analytical expression for the coupling constants is given in \cite{SM}. They all depend on  $\vecQ^{\sigma}$, meaning that the spin-spin interaction can be tuned electrically via the supercurrent. Since $\mathcal{H}_\text{eff}$ is obtained via perturbation theory, the expression is valid when the energy scale of the predicted effects is small compared to the terms in the original unperturbed Hamiltonian. We have confirmed numerically that the coupling constants are small compared to $\Delta_0, t$, and $\mu$. In $\mathcal{H}_\text{eff}$, $J_{i,j}$ and $K_{i,j}$ are exchange couplings, and the DM term $D_{i,j}$ is nonzero only in the presence of a spin supercurrent.  Importantly, $ L_{i,j}$ and $M_{i,j}$ enable the design of a spin lattice via supercurrents. They are given by
\begin{align}
L_{i,j} = \Upsilon\cos\left(\vecdQ R\right),\; M_{i,j} = -\Upsilon\sin\left(\vecdQ R\right),
\end{align}
where $\Upsilon\equiv \frac{\Lambda^2}{4N^2}\sum_{\vecq} A^{(2)}_{\vecq,\downarrow, \uparrow}\cos(\vecq\vectau)$, $\vecdQ = Q^\uparrow-Q^\downarrow$, $R =r_i+r_j$,  $\tau = r_j-r_i$, and the coefficients $A_{q,\downarrow,\uparrow}^{(2)}$ are large expressions that depend on energy eigenvalues and eigenstates of the superconducting Hamiltonian.
The dependence of $ L_{i,j}$ and $M_{i,j}$ on the absolute position $r_i+r_j$ stands in stark contrast to the conventional \textit{relative} position dependence $\tau$ that usually appears for RKKY interactions.
This is a remarkable feature that arises solely in the presence of a spin supercurrent. It is this dependence that enables the electrical design of a spin lattice, as we will show in the next section, and it can be understood physically as follows. A Cooper pair of spin $\uparrow$ electrons with one electron located at $r_i$ and the other electron at $r_j$ has the phase $Q^{\uparrow}R$. This phase is arbitrary since the physics cannot change under a U(1) gauge transformation of the order parameter. However, phase differences have physical consequences, and in the presence of a spin supercurrent, such a phase difference exists between a Cooper pair with spin $\uparrow$ and a Cooper pair with spin $\downarrow$. This phase difference is $\Delta Q^{\uparrow\downarrow}R$, which is precisely the quantity that the spin interaction coefficients depend on. The origin $r_i=0$ is defined by the relative phase $\Delta Q^{\uparrow\downarrow}$ of the superconducting condensates \cite{SM}.

We expect that the functional form of $\mathcal{H}_\text{eff}$ would remain unchanged in the strong coupling limit, for quantum spin systems, and if taking into account the local suppression of the magnitude of the OP, although the magnitude of the coupling constants would change. We also expect the this in the presence of spin-dependent terms in the normal state Hamiltionan, such as spin-orbit coupling or spin splitting, and if an intrinsic interaction between the magnetic adatoms is present in the first place. However, any such intrinsic exchange coupling would fall off extremely rapidly upon separating the impurity spins due to the lack of wavefunction overlap, leaving only the interaction mediated by the superconductor.
In the limit of large $S=|\vecS_i|$ and small coupling $\Lambda$ that we consider, the impurity spins may indeed be treated as classical and any Yu-Shiba-Rusinov (YSR) states would reside right at the gap edge
and provide a negligible spin screening \cite{balatsky_rmp_06} and negligible contribution to the spin-spin interaction \cite{deb_prb_21}. Screening currents induced from the impurity spins can be neglected for a  superconducting film of thickness $d\ll \lambda_L$ where $\lambda_L$ is the penetration depth \cite{meservey_physrep_94}, as is standard in the treatment of magnetic adatoms on superconducting surfaces  \cite{conte_arxiv_24}.

\begin{figure*}
    \includegraphics[]{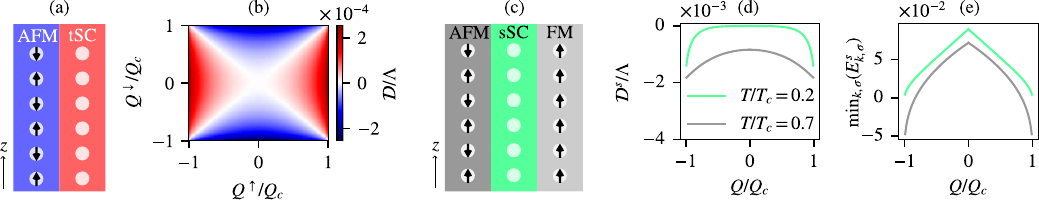}
    \caption{(a) A bilayer consisting of an antiferromagnetic insulator (AFM) and a triplet superconductor (tSC) with equal-spin Cooper pairs relative to the Néel order. (b) The supercurrent-induced contribution $\mathcal{D}$ to the magnon gap in the AFM in (a). The parameters used are $T=0.2T_c$ and $\mu=-t$, which gives $aQ_c\approx 0.039$. (c) A bilayer consisting of an AFM and a spin-split singlet superconductor (sSC). The spin splitting is induced by proximity to a ferromagnetic insulator (FM). The critical momentum is $aQ_c(T/T_c=0.2)\approx 0.043$ and $aQ_c(T/T_c=0.7)\approx 0.033$. (d)  The supercurrent-induced contribution $\mathcal{D}^s$ to the magnon gap in the AFM in (b). The spin splitting is $h=0.01t$ and the chemical potential is $\mu=-t$. (e) shows the corresponding minima of the quasiparticle energy dispersion. 
    }
    \label{fig:magnon}
\end{figure*}

\textit{Electrical tuning of the ground state} ---%
The ground state configuration is found by minimizing $\mathcal{H}_\text{eff}$ numerically. Note that $\mathcal{H}_\text{eff}$ is invariant under $\vecS_i \rightarrow -\vecS_i$, meaning that the ground state is always twofold degenerate. 
To illustrate how the ground state spin configuration is altered by a supercurrent, we consider two magnetic adatoms $\vecS_1$ and $\vecS_2$ placed on top of a superconductor with equal-spin Cooper pairs. The position of $\vecS_1$ is $r_1$, and the position of $\vecS_2$ is $r_1+\tau$. Figure \ref{fig:p-wave ground state}(a) shows the ground state of these two spins in the absence of a supercurrent for various $\tau$ values. The ground state configuration can be ferromagnetic or antiferromagnetic, depending on the distance between the spins. Upon applying a charge supercurrent, shown in Fig. \ref{fig:p-wave ground state}(b), the spin alignment axis changes for instance from the $xz$-plane to the $y$-axis for $\tau/a=3$. Since there is no spin current, the ground state configuration depends on the relative position of the spins $\tau$, but not on the absolute position of the spins. This changes when the supercurrent is spin-polarized. In Fig. \ref{fig:p-wave ground state}(c), a spin supercurrent flows through the superconductor, 
and we observe that the ground state configuration now depends on the absolute positions $r_1$ and $r_1+\tau$ of the spins in the lattice. This enables the possibility to design a lattice spin model that breaks translational invariance. The onset of non-zero $D_{i,j}$ and $M_{i,j}$ causes the spins to be non-collinear, and the angle difference from a collinear state is up to 65 degrees. Figure \ref{fig:p-wave ground state}(d) shows how the ground state can be tuned by a spin supercurrent. We have verified that all these properties are robust upon varying the chemical potential \cite{SM}. Moreover, we have verified that these properties also hold for a spin trimer consisting of three classical spins \cite{SM}. In Fig. \ref{fig:p-wave ground state}(e), we show how the trimer ground state varies with the spin supercurrent magnitude.

We note in passing that while the ground state configuration of the magnetic adatoms could be calculated non-perturbatively by diagonalizing numerically the Bogoliubov-de-Gennes Hamiltonian \cite{christensen2016spiral, pientka2013topological}, our perturbative approach allows us to extract an analytical effective spin-spin Hamiltonian in Eq. \eqref{eq:effective spin-spin model p-wave}. This determines the physical origin of the center-of-mass dependence as well as the precise analytical conditions for when the coupling constants are nonzero. Going beyond our perturbative approach to a regime where YSR states can reside deep within the gap could be an interesting future venue to explore emergent topological superconductivity by manipulating the state of a magnetic adatom lattice with supercurrents \cite{pientka2013topological,awoga2024controlling,li2016two}.

\textit{Magnon gap}---%
We now replace the two impurity spins considered up to now with an entire spin condensate that is proximity-coupled to a superconductor to demonstrate a new effect: electric control of a magnon gap in a spin-compensated magnet via a supercurrent.  Consider a bilayer consisting of an antiferromagnetic insulator and a superconductor with equal-spin Cooper pairs relative to the Néel order, as illustrated in Fig. \ref{fig:magnon}(a). We underline that our results apply to an altermagnetic insulator as well by generalizing the model to 2D \cite{SM}. The first order contribution to the magnon gap in the antiferromagnetic insulator is found by calculating $ \langle H_c\rangle_{\gamma}$ where $\gamma$ are the fermionic operators that diagonalize $H_\text{SC}$
, applying the Holstein-Primakoff transformation to the spin operators, and diagonalizing the full Hamiltonian for the antiferromagnetic insulator.
We find that $\langle H_c\rangle_{\gamma}$ is equivalent to an applied magnetic field, and the analytical supercurrent-induced contribution to the magnon gap is 
\begin{align}
        \mathcal{D} = -\frac{\Lambda}{2N} \sum_{k,\sigma} \sigma  \big( |u_{k,\sigma}|^2 n^+_{k,\sigma} + |v_{k,\sigma}|^2  n^-_{k,\sigma}  \big),
\end{align}
where $n^\pm_{k,\sigma} = n(\pm E_{\pm k+Q^{\sigma},\sigma})$ is the Fermi-Dirac distribution. The calculation details as well as definitions of the eigenenergy $E_{k,\sigma}$ and the coherence factors $u_{k,\sigma}$ and $v_{k,\sigma}$ are provided in \cite{SM}. 
 A pure charge supercurrent $(Q_\uparrow=Q_\downarrow)$ or a pure spin supercurrent $(Q_\uparrow=-Q_\downarrow)$ gives $\mathcal{D}=0$
 , while any combination of a charge and spin supercurrent $(Q_\uparrow \neq \pm Q_\downarrow)$ gives a tunable $\mathcal{D}\neq 0$. This is shown in Fig. \ref{fig:magnon}(b). In other words, a spin-polarized supercurrent tunes the magnitude of the magnon gap. The magnon gap is enhanced for one magnon species and suppressed for the other magnon species.

Interestingly, there may exist an even simpler way to control magnon gaps in antiferromagnetic insulators experimentally. Consider a bilayer consisting of a thin-film conventional $s$-wave superconductor with a spin-splitting field \cite{bergeret_rmp_18} and an antiferromagnetic insulator on top, as illustrated in Fig. \ref{fig:magnon}(c). The spin-splitting field in the superconductor can be achieved via an applied in-plane external field or by proximity to a ferromagnet when the thickness of the film is smaller than the London penetration depth. The induced spin splitting should be smaller than the critical value $h_c=\Delta_0/\sqrt{2}$ where $\Delta_0$ is the OP magnitude at $T=0$ \cite{clogston_prl_62, chandrasekhar_apl_62}. The supercurrent-induced change in the magnon gap is given by \cite{SM}:
\begin{align}
    {\mathcal{D}}^s= -\frac{\Lambda}{2N} \sum_{k,\sigma} \sigma n({E}^s_{k,\sigma}).
\end{align}
 The supercurrent-induced magnon gap against the Cooper pair momentum $Q$ is shown in Fig. \ref{fig:magnon}(d). The momentum is spin-independent because the Cooper pairs consist of electrons with opposite spins, and they cannot carry a spin current. Nevertheless, the charge supercurrent in this case controls the magnitude of the magnon gap. 
 The physics can be understood as follows. The spin splitting field $h$ induces a spin imbalance in the superconductor, which couples to the magnon density.
The spin imbalance is determined by the excitation spectrum 
in the superconductor, which is altered when applying a supercurrent. The gap $\text{min}_k (E^s_{k,\sigma})$ in the energy dispersion decreases for increasing $Q$, as shown in Fig. \ref{fig:magnon}(e). When the gap decreases, it becomes easier to excite quasiparticles, and their spin imbalance alters the magnon gap. This also explains why the supercurrent-induced magnon gap is larger at $T=0.7T_c$ compared to $T=0.2T_c$. In this way, electrical tuning of the magnon gap via a supercurrent is realized using conventional superconductivity. A possible material setup to measure this effect is a EuS/Nb bilayer with either an insulating antiferromagnet (Fe$_2$O$_3$) or altermagnet (MnTe) grown on top. A supercurrent flow can also influence the magnon gap and electron density of states in a superconductor/ferromagnet structure \cite{johnsen2021magnon, ianovskaia_prb_23}.

The fact that the intrinsic magnon gap in the antiferromagnetic insulator is enhanced by supercurrent for one magnon species but suppressed for the other 
lifts their degeneracy. 
Interestingly, this is the key to inducing magnon spin currents in antiferromagnets via for instance the spin Seebeck effect \cite{rezende_jap_19}. Our results thus demonstrate a route to electrically tunable magnon spin currents in antiferromagnetic insulators, controlled by a dissipationless supercurrent. With regard to the magnitude of the supercurrent-induced magnon gap, our approach is perturbative in nature and thus only small modifications of the magnon gap are within the regime of validity of our methodology. However, the physical mechanism, i.e., the supercurrent-induced spin polarization in the superconductor, is present even in the strong-coupling regime. Thus, a non-perturbative approach using stronger coupling would very likely increase this magnitude and leave the qualitative dependence of the magnon gap on supercurrent similar as that in the weak-coupling regime.

\begin{acknowledgments}
    A. Di Bernardo, C. Degen, S. K{\"o}lling, K. Knapp, U. Ognjanovic, M. Amundsen, and N. Banerjee are thanked for useful discussions. This work was supported by the Research Council of Norway through Grants No.\ 353894, 323766 and its Centres of Excellence funding scheme Grant No.\ 262633 ``QuSpin.''
    The numerical calculations were performed on resources provided by Sigma2---the National Infrastructure for High Performance Computing and Data Storage in Norway, project NN9577K. 
\end{acknowledgments}

\bibliography{references}

\begin{thebibliography}{65}%
\makeatletter
\providecommand \@ifxundefined [1]{%
 \@ifx{#1\undefined}
}%
\providecommand \@ifnum [1]{%
 \ifnum #1\expandafter \@firstoftwo
 \else \expandafter \@secondoftwo
 \fi
}%
\providecommand \@ifx [1]{%
 \ifx #1\expandafter \@firstoftwo
 \else \expandafter \@secondoftwo
 \fi
}%
\providecommand \natexlab [1]{#1}%
\providecommand \enquote  [1]{``#1''}%
\providecommand \bibnamefont  [1]{#1}%
\providecommand \bibfnamefont [1]{#1}%
\providecommand \citenamefont [1]{#1}%
\providecommand \href@noop [0]{\@secondoftwo}%
\providecommand \href [0]{\begingroup \@sanitize@url \@href}%
\providecommand \@href[1]{\@@startlink{#1}\@@href}%
\providecommand \@@href[1]{\endgroup#1\@@endlink}%
\providecommand \@sanitize@url [0]{\catcode `\\12\catcode `\$12\catcode `\&12\catcode `\#12\catcode `\^12\catcode `\_12\catcode `\%12\relax}%
\providecommand \@@startlink[1]{}%
\providecommand \@@endlink[0]{}%
\providecommand \url  [0]{\begingroup\@sanitize@url \@url }%
\providecommand \@url [1]{\endgroup\@href {#1}{\urlprefix }}%
\providecommand \urlprefix  [0]{URL }%
\providecommand \Eprint [0]{\href }%
\providecommand \doibase [0]{http://dx.doi.org/}%
\providecommand \selectlanguage [0]{\@gobble}%
\providecommand \bibinfo  [0]{\@secondoftwo}%
\providecommand \bibfield  [0]{\@secondoftwo}%
\providecommand \translation [1]{[#1]}%
\providecommand \BibitemOpen [0]{}%
\providecommand \bibitemStop [0]{}%
\providecommand \bibitemNoStop [0]{.\EOS\space}%
\providecommand \EOS [0]{\spacefactor3000\relax}%
\providecommand \BibitemShut  [1]{\csname bibitem#1\endcsname}%
\let\auto@bib@innerbib\@empty
\bibitem [{\citenamefont {Roberts}\ \emph {et~al.}(2025)\citenamefont {Roberts}, \citenamefont {Abudayyeh}, \citenamefont {Li},\ and\ \citenamefont {Li}}]{roberts2025quantum}%
  \BibitemOpen
  \bibfield  {author} {\bibinfo {author} {\bibfnamefont {H.}~\bibnamefont {Roberts}}, \bibinfo {author} {\bibfnamefont {H.}~\bibnamefont {Abudayyeh}}, \bibinfo {author} {\bibfnamefont {X.}~\bibnamefont {Li}}, \ and\ \bibinfo {author} {\bibfnamefont {X.}~\bibnamefont {Li}},\ }\bibfield  {title} {\enquote {\bibinfo {title} {{Quantum Sensing with Spin Defects Beyond Diamond}},}\ }\href {https://pubs.acs.org/doi/full/10.1021/acsnano.5c00802} {\bibfield  {journal} {\bibinfo  {journal} {ACS Nano}\ } (\bibinfo {year} {2025})}\BibitemShut {NoStop}%
\bibitem [{\citenamefont {Fang}\ \emph {et~al.}(2024)\citenamefont {Fang}, \citenamefont {Wang}, \citenamefont {Marie},\ and\ \citenamefont {Sun}}]{fang2024quantum}%
  \BibitemOpen
  \bibfield  {author} {\bibinfo {author} {\bibfnamefont {H.-H.}\ \bibnamefont {Fang}}, \bibinfo {author} {\bibfnamefont {X.-J.}\ \bibnamefont {Wang}}, \bibinfo {author} {\bibfnamefont {X.}~\bibnamefont {Marie}}, \ and\ \bibinfo {author} {\bibfnamefont {H.-B.}\ \bibnamefont {Sun}},\ }\bibfield  {title} {\enquote {\bibinfo {title} {{Quantum sensing with optically accessible spin defects in van der Waals layered materials}},}\ }\href {https://www.nature.com/articles/s41377-024-01630-y} {\bibfield  {journal} {\bibinfo  {journal} {Light: Science \& Applications}\ }\textbf {\bibinfo {volume} {13}},\ \bibinfo {pages} {303} (\bibinfo {year} {2024})}\BibitemShut {NoStop}%
\bibitem [{\citenamefont {Tanamoto}\ and\ \citenamefont {Ono}(2021)}]{tanamoto2021compact}%
  \BibitemOpen
  \bibfield  {author} {\bibinfo {author} {\bibfnamefont {T.}~\bibnamefont {Tanamoto}}\ and\ \bibinfo {author} {\bibfnamefont {K.}~\bibnamefont {Ono}},\ }\bibfield  {title} {\enquote {\bibinfo {title} {Compact spin qubits using the common gate structure of fin field-effect transistors},}\ }\href {https://pubs.aip.org/aip/adv/article/11/4/045004/976522} {\bibfield  {journal} {\bibinfo  {journal} {AIP Advances}\ }\textbf {\bibinfo {volume} {11}} (\bibinfo {year} {2021})}\BibitemShut {NoStop}%
\bibitem [{\citenamefont {Yang}\ \emph {et~al.}(2016)\citenamefont {Yang}, \citenamefont {Hsu}, \citenamefont {Stano}, \citenamefont {Klinovaja},\ and\ \citenamefont {Loss}}]{yang2016long}%
  \BibitemOpen
  \bibfield  {author} {\bibinfo {author} {\bibfnamefont {G.}~\bibnamefont {Yang}}, \bibinfo {author} {\bibfnamefont {C.-H.}\ \bibnamefont {Hsu}}, \bibinfo {author} {\bibfnamefont {P.}~\bibnamefont {Stano}}, \bibinfo {author} {\bibfnamefont {J.}~\bibnamefont {Klinovaja}}, \ and\ \bibinfo {author} {\bibfnamefont {D.}~\bibnamefont {Loss}},\ }\bibfield  {title} {\enquote {\bibinfo {title} {{Long-distance entanglement of spin qubits via quantum Hall edge states}},}\ }\href {https://journals.aps.org/prb/abstract/10.1103/PhysRevB.93.075301} {\bibfield  {journal} {\bibinfo  {journal} {Physical Review B}\ }\textbf {\bibinfo {volume} {93}},\ \bibinfo {pages} {075301} (\bibinfo {year} {2016})}\BibitemShut {NoStop}%
\bibitem [{\citenamefont {Mishra}\ \emph {et~al.}(2021)\citenamefont {Mishra}, \citenamefont {Simon}, \citenamefont {Hyart},\ and\ \citenamefont {Trif}}]{mishra2021yu}%
  \BibitemOpen
  \bibfield  {author} {\bibinfo {author} {\bibfnamefont {A.}~\bibnamefont {Mishra}}, \bibinfo {author} {\bibfnamefont {P.}~\bibnamefont {Simon}}, \bibinfo {author} {\bibfnamefont {T.}~\bibnamefont {Hyart}}, \ and\ \bibinfo {author} {\bibfnamefont {M.}~\bibnamefont {Trif}},\ }\bibfield  {title} {\enquote {\bibinfo {title} {{Yu-Shiba-Rusinov qubit}},}\ }\href {https://journals.aps.org/prxquantum/abstract/10.1103/PRXQuantum.2.040347} {\bibfield  {journal} {\bibinfo  {journal} {PRX Quantum}\ }\textbf {\bibinfo {volume} {2}},\ \bibinfo {pages} {040347} (\bibinfo {year} {2021})}\BibitemShut {NoStop}%
\bibitem [{\citenamefont {Zhang}\ \emph {et~al.}(2024)\citenamefont {Zhang}, \citenamefont {Sun}, \citenamefont {Cao}, \citenamefont {Yang}, \citenamefont {Yang}, \citenamefont {Lu}, \citenamefont {Du}, \citenamefont {Hu}, \citenamefont {Feng}, \citenamefont {Wang}, \citenamefont {Cai}, \citenamefont {Cui}, \citenamefont {Piao}, \citenamefont {Zhao},\ and\ \citenamefont {Zhao}}]{zhang_sciadv_24}%
  \BibitemOpen
  \bibfield  {author} {\bibinfo {author} {\bibfnamefont {Yike}\ \bibnamefont {Zhang}}, \bibinfo {author} {\bibfnamefont {Weideng}\ \bibnamefont {Sun}}, \bibinfo {author} {\bibfnamefont {Kaihua}\ \bibnamefont {Cao}}, \bibinfo {author} {\bibfnamefont {Xiao-Xue}\ \bibnamefont {Yang}}, \bibinfo {author} {\bibfnamefont {Yongqiang}\ \bibnamefont {Yang}}, \bibinfo {author} {\bibfnamefont {Shiyang}\ \bibnamefont {Lu}}, \bibinfo {author} {\bibfnamefont {Ao}~\bibnamefont {Du}}, \bibinfo {author} {\bibfnamefont {Chaoqun}\ \bibnamefont {Hu}}, \bibinfo {author} {\bibfnamefont {Ce}~\bibnamefont {Feng}}, \bibinfo {author} {\bibfnamefont {Yutong}\ \bibnamefont {Wang}}, \bibinfo {author} {\bibfnamefont {Jianwang}\ \bibnamefont {Cai}}, \bibinfo {author} {\bibfnamefont {Baoshan}\ \bibnamefont {Cui}}, \bibinfo {author} {\bibfnamefont {Hong-Guang}\ \bibnamefont {Piao}}, \bibinfo {author} {\bibfnamefont {Weisheng}\ \bibnamefont {Zhao}}, \ and\ \bibinfo {author} {\bibfnamefont {Yonggang}\ \bibnamefont {Zhao}},\ }\bibfield  {title}
  {\enquote {\bibinfo {title} {Electric-field control of nonvolatile resistance state of perpendicular magnetic tunnel junction via magnetoelectric coupling},}\ }\href {\doibase 10.1126/sciadv.adl4633} {\bibfield  {journal} {\bibinfo  {journal} {Science Advances}\ }\textbf {\bibinfo {volume} {10}} (\bibinfo {year} {2024}),\ 10.1126/sciadv.adl4633}\BibitemShut {NoStop}%
\bibitem [{\citenamefont {Nadj-Perge}\ \emph {et~al.}(2014)\citenamefont {Nadj-Perge}, \citenamefont {Drozdov}, \citenamefont {Li}, \citenamefont {Chen}, \citenamefont {Jeon}, \citenamefont {Seo}, \citenamefont {MacDonald}, \citenamefont {Bernevig},\ and\ \citenamefont {Yazdani}}]{nadj2014observation}%
  \BibitemOpen
  \bibfield  {author} {\bibinfo {author} {\bibfnamefont {S.}~\bibnamefont {Nadj-Perge}}, \bibinfo {author} {\bibfnamefont {I.~K.}\ \bibnamefont {Drozdov}}, \bibinfo {author} {\bibfnamefont {J.}~\bibnamefont {Li}}, \bibinfo {author} {\bibfnamefont {H.}~\bibnamefont {Chen}}, \bibinfo {author} {\bibfnamefont {S.}~\bibnamefont {Jeon}}, \bibinfo {author} {\bibfnamefont {J.}~\bibnamefont {Seo}}, \bibinfo {author} {\bibfnamefont {A.~H.}\ \bibnamefont {MacDonald}}, \bibinfo {author} {\bibfnamefont {B.~A.}\ \bibnamefont {Bernevig}}, \ and\ \bibinfo {author} {\bibfnamefont {A.}~\bibnamefont {Yazdani}},\ }\bibfield  {title} {\enquote {\bibinfo {title} {{Observation of Majorana fermions in ferromagnetic atomic chains on a superconductor}},}\ }\href {https://www.science.org/doi/full/10.1126/science.1259327} {\bibfield  {journal} {\bibinfo  {journal} {Science}\ }\textbf {\bibinfo {volume} {346}},\ \bibinfo {pages} {602--607} (\bibinfo {year} {2014})}\BibitemShut {NoStop}%
\bibitem [{\citenamefont {Pupim}\ and\ \citenamefont {Scheurer}(2025)}]{pupim2025adatom}%
  \BibitemOpen
  \bibfield  {author} {\bibinfo {author} {\bibfnamefont {L.~V.}\ \bibnamefont {Pupim}}\ and\ \bibinfo {author} {\bibfnamefont {M.~S.}\ \bibnamefont {Scheurer}},\ }\bibfield  {title} {\enquote {\bibinfo {title} {Adatom engineering magnetic order in superconductors: Applications to altermagnetic superconductivity},}\ }\href {https://journals.aps.org/prl/abstract/10.1103/PhysRevLett.134.146001} {\bibfield  {journal} {\bibinfo  {journal} {Physical Review Letters}\ }\textbf {\bibinfo {volume} {134}},\ \bibinfo {pages} {146001} (\bibinfo {year} {2025})}\BibitemShut {NoStop}%
\bibitem [{\citenamefont {Kim}\ \emph {et~al.}(2018)\citenamefont {Kim}, \citenamefont {Palacio-Morales}, \citenamefont {Posske}, \citenamefont {R{\'o}zsa}, \citenamefont {Palot{\'a}s}, \citenamefont {Szunyogh}, \citenamefont {Thorwart},\ and\ \citenamefont {Wiesendanger}}]{kim2018toward}%
  \BibitemOpen
  \bibfield  {author} {\bibinfo {author} {\bibfnamefont {H.}~\bibnamefont {Kim}}, \bibinfo {author} {\bibfnamefont {A.}~\bibnamefont {Palacio-Morales}}, \bibinfo {author} {\bibfnamefont {T.}~\bibnamefont {Posske}}, \bibinfo {author} {\bibfnamefont {L.}~\bibnamefont {R{\'o}zsa}}, \bibinfo {author} {\bibfnamefont {K.}~\bibnamefont {Palot{\'a}s}}, \bibinfo {author} {\bibfnamefont {L.}~\bibnamefont {Szunyogh}}, \bibinfo {author} {\bibfnamefont {M.}~\bibnamefont {Thorwart}}, \ and\ \bibinfo {author} {\bibfnamefont {R.}~\bibnamefont {Wiesendanger}},\ }\bibfield  {title} {\enquote {\bibinfo {title} {{Toward tailoring Majorana bound states in artificially constructed magnetic atom chains on elemental superconductors}},}\ }\href {https://www.science.org/doi/full/10.1126/sciadv.aar5251} {\bibfield  {journal} {\bibinfo  {journal} {Science Advances}\ }\textbf {\bibinfo {volume} {4}},\ \bibinfo {pages} {eaar5251} (\bibinfo {year} {2018})}\BibitemShut {NoStop}%
\bibitem [{\citenamefont {Heimes}\ \emph {et~al.}(2015)\citenamefont {Heimes}, \citenamefont {Mendler},\ and\ \citenamefont {Kotetes}}]{heimes2015interplay}%
  \BibitemOpen
  \bibfield  {author} {\bibinfo {author} {\bibfnamefont {A.}~\bibnamefont {Heimes}}, \bibinfo {author} {\bibfnamefont {D.}~\bibnamefont {Mendler}}, \ and\ \bibinfo {author} {\bibfnamefont {P.}~\bibnamefont {Kotetes}},\ }\bibfield  {title} {\enquote {\bibinfo {title} {{Interplay of topological phases in magnetic adatom-chains on top of a Rashba superconducting surface}},}\ }\href {https://iopscience.iop.org/article/10.1088/1367-2630/17/2/023051/meta} {\bibfield  {journal} {\bibinfo  {journal} {New Journal of Physics}\ }\textbf {\bibinfo {volume} {17}},\ \bibinfo {pages} {023051} (\bibinfo {year} {2015})}\BibitemShut {NoStop}%
\bibitem [{\citenamefont {Mohanta}\ \emph {et~al.}(2018)\citenamefont {Mohanta}, \citenamefont {Kampf},\ and\ \citenamefont {Kopp}}]{mohanta2018supercurrent}%
  \BibitemOpen
  \bibfield  {author} {\bibinfo {author} {\bibfnamefont {N.}~\bibnamefont {Mohanta}}, \bibinfo {author} {\bibfnamefont {A.~P.}\ \bibnamefont {Kampf}}, \ and\ \bibinfo {author} {\bibfnamefont {T.}~\bibnamefont {Kopp}},\ }\bibfield  {title} {\enquote {\bibinfo {title} {Supercurrent as a probe for topological superconductivity in magnetic adatom chains},}\ }\href {https://journals.aps.org/prb/abstract/10.1103/PhysRevB.97.214507} {\bibfield  {journal} {\bibinfo  {journal} {Physical Review B}\ }\textbf {\bibinfo {volume} {97}},\ \bibinfo {pages} {214507} (\bibinfo {year} {2018})}\BibitemShut {NoStop}%
\bibitem [{\citenamefont {Yazdani}\ \emph {et~al.}(1997)\citenamefont {Yazdani}, \citenamefont {Jones}, \citenamefont {Lutz}, \citenamefont {Crommie},\ and\ \citenamefont {Eigler}}]{yazdani1997probing}%
  \BibitemOpen
  \bibfield  {author} {\bibinfo {author} {\bibfnamefont {A.}~\bibnamefont {Yazdani}}, \bibinfo {author} {\bibfnamefont {B.~A.}\ \bibnamefont {Jones}}, \bibinfo {author} {\bibfnamefont {C.~P.}\ \bibnamefont {Lutz}}, \bibinfo {author} {\bibfnamefont {M.~F.}\ \bibnamefont {Crommie}}, \ and\ \bibinfo {author} {\bibfnamefont {D.~M.}\ \bibnamefont {Eigler}},\ }\bibfield  {title} {\enquote {\bibinfo {title} {Probing the local effects of magnetic impurities on superconductivity},}\ }\href {https://www.science.org/doi/full/10.1126/science.275.5307.1767} {\bibfield  {journal} {\bibinfo  {journal} {Science}\ }\textbf {\bibinfo {volume} {275}},\ \bibinfo {pages} {1767--1770} (\bibinfo {year} {1997})}\BibitemShut {NoStop}%
\bibitem [{\citenamefont {R{\"o}ntynen}\ and\ \citenamefont {Ojanen}(2015)}]{rontynen2015topological}%
  \BibitemOpen
  \bibfield  {author} {\bibinfo {author} {\bibfnamefont {J.}~\bibnamefont {R{\"o}ntynen}}\ and\ \bibinfo {author} {\bibfnamefont {T.}~\bibnamefont {Ojanen}},\ }\bibfield  {title} {\enquote {\bibinfo {title} {{Topological superconductivity and high Chern numbers in 2D ferromagnetic Shiba lattices}},}\ }\href {https://journals.aps.org/prl/abstract/10.1103/PhysRevLett.114.236803} {\bibfield  {journal} {\bibinfo  {journal} {Physical Review Letters}\ }\textbf {\bibinfo {volume} {114}},\ \bibinfo {pages} {236803} (\bibinfo {year} {2015})}\BibitemShut {NoStop}%
\bibitem [{\citenamefont {Kimura}(2007)}]{kimura2007spiral}%
  \BibitemOpen
  \bibfield  {author} {\bibinfo {author} {\bibfnamefont {T.}~\bibnamefont {Kimura}},\ }\bibfield  {title} {\enquote {\bibinfo {title} {Spiral magnets as magnetoelectrics},}\ }\href {https://www.annualreviews.org/content/journals/10.1146/annurev.matsci.37.052506.084259} {\bibfield  {journal} {\bibinfo  {journal} {Annu. Rev. Mater. Res.}\ }\textbf {\bibinfo {volume} {37}},\ \bibinfo {pages} {387--413} (\bibinfo {year} {2007})}\BibitemShut {NoStop}%
\bibitem [{\citenamefont {Tokura}\ and\ \citenamefont {Seki}(2010)}]{tokura2010multiferroics}%
  \BibitemOpen
  \bibfield  {author} {\bibinfo {author} {\bibfnamefont {Y.}~\bibnamefont {Tokura}}\ and\ \bibinfo {author} {\bibfnamefont {S.}~\bibnamefont {Seki}},\ }\bibfield  {title} {\enquote {\bibinfo {title} {Multiferroics with spiral spin orders},}\ }\href {https://advanced.onlinelibrary.wiley.com/doi/full/10.1002/adma.200901961} {\bibfield  {journal} {\bibinfo  {journal} {Advanced Materials}\ }\textbf {\bibinfo {volume} {22}},\ \bibinfo {pages} {1554--1565} (\bibinfo {year} {2010})}\BibitemShut {NoStop}%
\bibitem [{\citenamefont {Nagaosa}\ and\ \citenamefont {Tokura}(2013)}]{nagaosa2013topological}%
  \BibitemOpen
  \bibfield  {author} {\bibinfo {author} {\bibfnamefont {N.}~\bibnamefont {Nagaosa}}\ and\ \bibinfo {author} {\bibfnamefont {Y.}~\bibnamefont {Tokura}},\ }\bibfield  {title} {\enquote {\bibinfo {title} {{Topological properties and dynamics of magnetic skyrmions}},}\ }\href {https://www.nature.com/articles/nnano.2013.243} {\bibfield  {journal} {\bibinfo  {journal} {Nature Nanotechnology}\ }\textbf {\bibinfo {volume} {8}},\ \bibinfo {pages} {899--911} (\bibinfo {year} {2013})}\BibitemShut {NoStop}%
\bibitem [{\citenamefont {Zhang}\ \emph {et~al.}(2020)\citenamefont {Zhang}, \citenamefont {Zhou}, \citenamefont {Song}, \citenamefont {Park}, \citenamefont {Xia}, \citenamefont {Ezawa}, \citenamefont {Liu}, \citenamefont {Zhao}, \citenamefont {Zhao},\ and\ \citenamefont {Woo}}]{zhang2020skyrmion}%
  \BibitemOpen
  \bibfield  {author} {\bibinfo {author} {\bibfnamefont {X.}~\bibnamefont {Zhang}}, \bibinfo {author} {\bibfnamefont {Y.}~\bibnamefont {Zhou}}, \bibinfo {author} {\bibfnamefont {K.~M.}\ \bibnamefont {Song}}, \bibinfo {author} {\bibfnamefont {T.-E.}\ \bibnamefont {Park}}, \bibinfo {author} {\bibfnamefont {J.}~\bibnamefont {Xia}}, \bibinfo {author} {\bibfnamefont {M.}~\bibnamefont {Ezawa}}, \bibinfo {author} {\bibfnamefont {X.}~\bibnamefont {Liu}}, \bibinfo {author} {\bibfnamefont {W.}~\bibnamefont {Zhao}}, \bibinfo {author} {\bibfnamefont {G.}~\bibnamefont {Zhao}}, \ and\ \bibinfo {author} {\bibfnamefont {S.}~\bibnamefont {Woo}},\ }\bibfield  {title} {\enquote {\bibinfo {title} {{Skyrmion-electronics: writing, deleting, reading and processing magnetic skyrmions toward spintronic applications}},}\ }\href {https://iopscience.iop.org/article/10.1088/1361-648X/ab5488/meta} {\bibfield  {journal} {\bibinfo  {journal} {Journal of Physics: Condensed Matter}\ }\textbf {\bibinfo {volume} {32}},\ \bibinfo {pages}
  {143001} (\bibinfo {year} {2020})}\BibitemShut {NoStop}%
\bibitem [{\citenamefont {Marrows}\ and\ \citenamefont {Zeissler}(2021)}]{marrows2021perspective}%
  \BibitemOpen
  \bibfield  {author} {\bibinfo {author} {\bibfnamefont {C.~H.}\ \bibnamefont {Marrows}}\ and\ \bibinfo {author} {\bibfnamefont {K.}~\bibnamefont {Zeissler}},\ }\bibfield  {title} {\enquote {\bibinfo {title} {{Perspective on skyrmion spintronics}},}\ }\href {https://pubs.aip.org/aip/apl/article/119/25/250502/40665} {\bibfield  {journal} {\bibinfo  {journal} {Applied Physics Letters}\ }\textbf {\bibinfo {volume} {119}} (\bibinfo {year} {2021})}\BibitemShut {NoStop}%
\bibitem [{\citenamefont {Rimmler}\ \emph {et~al.}(2025)\citenamefont {Rimmler}, \citenamefont {Pal},\ and\ \citenamefont {Parkin}}]{rimmler2025non-collinear}%
  \BibitemOpen
  \bibfield  {author} {\bibinfo {author} {\bibfnamefont {B.~H.}\ \bibnamefont {Rimmler}}, \bibinfo {author} {\bibfnamefont {B.}~\bibnamefont {Pal}}, \ and\ \bibinfo {author} {\bibfnamefont {S.~S.~P.}\ \bibnamefont {Parkin}},\ }\bibfield  {title} {\enquote {\bibinfo {title} {{Non-collinear antiferromagnetic spintronics}},}\ }\href {https://www.nature.com/articles/s41578-024-00706-w} {\bibfield  {journal} {\bibinfo  {journal} {Nature Reviews Materials}\ }\textbf {\bibinfo {volume} {10}},\ \bibinfo {pages} {109--127} (\bibinfo {year} {2025})}\BibitemShut {NoStop}%
\bibitem [{\citenamefont {Soumyanarayanan}\ \emph {et~al.}(2016)\citenamefont {Soumyanarayanan}, \citenamefont {Reyren}, \citenamefont {Fert},\ and\ \citenamefont {Panagopoulos}}]{soumyanarayanan2016emergent}%
  \BibitemOpen
  \bibfield  {author} {\bibinfo {author} {\bibfnamefont {A.}~\bibnamefont {Soumyanarayanan}}, \bibinfo {author} {\bibfnamefont {N.}~\bibnamefont {Reyren}}, \bibinfo {author} {\bibfnamefont {A.}~\bibnamefont {Fert}}, \ and\ \bibinfo {author} {\bibfnamefont {C.}~\bibnamefont {Panagopoulos}},\ }\bibfield  {title} {\enquote {\bibinfo {title} {{Emergent phenomena induced by spin--orbit coupling at surfaces and interfaces}},}\ }\href {https://www.nature.com/articles/nature19820} {\bibfield  {journal} {\bibinfo  {journal} {Nature}\ }\textbf {\bibinfo {volume} {539}},\ \bibinfo {pages} {509--517} (\bibinfo {year} {2016})}\BibitemShut {NoStop}%
\bibitem [{\citenamefont {Ouassou}\ \emph {et~al.}(2025)\citenamefont {Ouassou}, \citenamefont {Yokoyama},\ and\ \citenamefont {Linder}}]{ouassou2025dzyaloshinskii}%
  \BibitemOpen
  \bibfield  {author} {\bibinfo {author} {\bibfnamefont {J.~A.}\ \bibnamefont {Ouassou}}, \bibinfo {author} {\bibfnamefont {T.}~\bibnamefont {Yokoyama}}, \ and\ \bibinfo {author} {\bibfnamefont {J.}~\bibnamefont {Linder}},\ }\bibfield  {title} {\enquote {\bibinfo {title} {{Dzyaloshinskii-Moriya-type spin-spin interaction from mixed-parity superconductivity}},}\ }\href {https://journals.aps.org/prb/abstract/10.1103/PhysRevB.111.L060504} {\bibfield  {journal} {\bibinfo  {journal} {Physical Review B}\ }\textbf {\bibinfo {volume} {111}},\ \bibinfo {pages} {L060504} (\bibinfo {year} {2025})}\BibitemShut {NoStop}%
\bibitem [{\citenamefont {Chumak}\ \emph {et~al.}(2015)\citenamefont {Chumak}, \citenamefont {Vasyuchka}, \citenamefont {Serga},\ and\ \citenamefont {Hillebrands}}]{chumak2015magnon}%
  \BibitemOpen
  \bibfield  {author} {\bibinfo {author} {\bibfnamefont {A.~V.}\ \bibnamefont {Chumak}}, \bibinfo {author} {\bibfnamefont {V.~I.}\ \bibnamefont {Vasyuchka}}, \bibinfo {author} {\bibfnamefont {A.~A.}\ \bibnamefont {Serga}}, \ and\ \bibinfo {author} {\bibfnamefont {B.}~\bibnamefont {Hillebrands}},\ }\bibfield  {title} {\enquote {\bibinfo {title} {Magnon spintronics},}\ }\href {https://www.nature.com/articles/nphys3347} {\bibfield  {journal} {\bibinfo  {journal} {Nature Physics}\ }\textbf {\bibinfo {volume} {11}},\ \bibinfo {pages} {453--461} (\bibinfo {year} {2015})}\BibitemShut {NoStop}%
\bibitem [{\citenamefont {Oba}\ \emph {et~al.}(2015)\citenamefont {Oba}, \citenamefont {Nakamura}, \citenamefont {Akiyama}, \citenamefont {Ito}, \citenamefont {Weinert},\ and\ \citenamefont {Freeman}}]{oba2015electric}%
  \BibitemOpen
  \bibfield  {author} {\bibinfo {author} {\bibfnamefont {M.}~\bibnamefont {Oba}}, \bibinfo {author} {\bibfnamefont {K.}~\bibnamefont {Nakamura}}, \bibinfo {author} {\bibfnamefont {T.}~\bibnamefont {Akiyama}}, \bibinfo {author} {\bibfnamefont {T.}~\bibnamefont {Ito}}, \bibinfo {author} {\bibfnamefont {M.}~\bibnamefont {Weinert}}, \ and\ \bibinfo {author} {\bibfnamefont {A.~J.}\ \bibnamefont {Freeman}},\ }\bibfield  {title} {\enquote {\bibinfo {title} {{Electric-field-induced modification of the magnon energy, exchange interaction, and Curie temperature of transition-metal thin films}},}\ }\href {https://journals.aps.org/prl/abstract/10.1103/PhysRevLett.114.107202} {\bibfield  {journal} {\bibinfo  {journal} {Physical Review Letters}\ }\textbf {\bibinfo {volume} {114}},\ \bibinfo {pages} {107202} (\bibinfo {year} {2015})}\BibitemShut {NoStop}%
\bibitem [{\citenamefont {Pradipto}\ \emph {et~al.}(2017)\citenamefont {Pradipto}, \citenamefont {Akiyama}, \citenamefont {Ito},\ and\ \citenamefont {Nakamura}}]{pradipto2017mechanism}%
  \BibitemOpen
  \bibfield  {author} {\bibinfo {author} {\bibfnamefont {A.-M.}\ \bibnamefont {Pradipto}}, \bibinfo {author} {\bibfnamefont {T.}~\bibnamefont {Akiyama}}, \bibinfo {author} {\bibfnamefont {T.}~\bibnamefont {Ito}}, \ and\ \bibinfo {author} {\bibfnamefont {K.}~\bibnamefont {Nakamura}},\ }\bibfield  {title} {\enquote {\bibinfo {title} {{Mechanism and electric field induced modification of magnetic exchange stiffness in transition metal thin films on MgO (001)}},}\ }\href {https://journals.aps.org/prb/abstract/10.1103/PhysRevB.96.014425} {\bibfield  {journal} {\bibinfo  {journal} {Physical Review B}\ }\textbf {\bibinfo {volume} {96}},\ \bibinfo {pages} {014425} (\bibinfo {year} {2017})}\BibitemShut {NoStop}%
\bibitem [{\citenamefont {Yoshii}\ \emph {et~al.}(2020)\citenamefont {Yoshii}, \citenamefont {Ohshima}, \citenamefont {Ando}, \citenamefont {Shinjo},\ and\ \citenamefont {Shiraishi}}]{yoshii2020detection}%
  \BibitemOpen
  \bibfield  {author} {\bibinfo {author} {\bibfnamefont {S.}~\bibnamefont {Yoshii}}, \bibinfo {author} {\bibfnamefont {R.}~\bibnamefont {Ohshima}}, \bibinfo {author} {\bibfnamefont {Y.}~\bibnamefont {Ando}}, \bibinfo {author} {\bibfnamefont {T.}~\bibnamefont {Shinjo}}, \ and\ \bibinfo {author} {\bibfnamefont {M.}~\bibnamefont {Shiraishi}},\ }\bibfield  {title} {\enquote {\bibinfo {title} {{Detection of ferromagnetic resonance from 1 nm-thick Co}},}\ }\href {https://www.nature.com/articles/s41598-020-72760-7} {\bibfield  {journal} {\bibinfo  {journal} {Scientific Reports}\ }\textbf {\bibinfo {volume} {10}},\ \bibinfo {pages} {15764} (\bibinfo {year} {2020})}\BibitemShut {NoStop}%
\bibitem [{\citenamefont {Zhu}\ \emph {et~al.}(2021)\citenamefont {Zhu}, \citenamefont {Zhang}, \citenamefont {Wang}, \citenamefont {Dos~Santos}, \citenamefont {Song}, \citenamefont {Mueller}, \citenamefont {Schmalzl}, \citenamefont {Schmidt}, \citenamefont {Ivanov}, \citenamefont {Park} \emph {et~al.}}]{zhu2021topological}%
  \BibitemOpen
  \bibfield  {author} {\bibinfo {author} {\bibfnamefont {F.}~\bibnamefont {Zhu}}, \bibinfo {author} {\bibfnamefont {L.}~\bibnamefont {Zhang}}, \bibinfo {author} {\bibfnamefont {X.}~\bibnamefont {Wang}}, \bibinfo {author} {\bibfnamefont {F.~J.}\ \bibnamefont {Dos~Santos}}, \bibinfo {author} {\bibfnamefont {J.}~\bibnamefont {Song}}, \bibinfo {author} {\bibfnamefont {T.}~\bibnamefont {Mueller}}, \bibinfo {author} {\bibfnamefont {K.}~\bibnamefont {Schmalzl}}, \bibinfo {author} {\bibfnamefont {W.~F.}\ \bibnamefont {Schmidt}}, \bibinfo {author} {\bibfnamefont {A.}~\bibnamefont {Ivanov}}, \bibinfo {author} {\bibfnamefont {J.~T.}\ \bibnamefont {Park}},  \emph {et~al.},\ }\bibfield  {title} {\enquote {\bibinfo {title} {{Topological magnon insulators in two-dimensional van der Waals ferromagnets CrSiTe$_3$ and CrGeTe$_3$: Toward intrinsic gap-tunability}},}\ }\href {https://www.science.org/doi/full/10.1126/sciadv.abi7532} {\bibfield  {journal} {\bibinfo  {journal} {Science Advances}\ }\textbf {\bibinfo {volume} {7}},\
  \bibinfo {pages} {eabi7532} (\bibinfo {year} {2021})}\BibitemShut {NoStop}%
\bibitem [{\citenamefont {Yumnam}\ \emph {et~al.}(2024)\citenamefont {Yumnam}, \citenamefont {Moseley}, \citenamefont {Paddison}, \citenamefont {Suggs}, \citenamefont {Zappala}, \citenamefont {Parker}, \citenamefont {Granroth}, \citenamefont {Morris}, \citenamefont {Polash}, \citenamefont {Vashaee} \emph {et~al.}}]{yumnam2024magnon}%
  \BibitemOpen
  \bibfield  {author} {\bibinfo {author} {\bibfnamefont {G.}~\bibnamefont {Yumnam}}, \bibinfo {author} {\bibfnamefont {D.~H.}\ \bibnamefont {Moseley}}, \bibinfo {author} {\bibfnamefont {J.~A.~M.}\ \bibnamefont {Paddison}}, \bibinfo {author} {\bibfnamefont {C.~Z.}\ \bibnamefont {Suggs}}, \bibinfo {author} {\bibfnamefont {E.}~\bibnamefont {Zappala}}, \bibinfo {author} {\bibfnamefont {D.~S.}\ \bibnamefont {Parker}}, \bibinfo {author} {\bibfnamefont {G.~E.}\ \bibnamefont {Granroth}}, \bibinfo {author} {\bibfnamefont {G.~D.}\ \bibnamefont {Morris}}, \bibinfo {author} {\bibfnamefont {M.~M.~H.}\ \bibnamefont {Polash}}, \bibinfo {author} {\bibfnamefont {D.}~\bibnamefont {Vashaee}},  \emph {et~al.},\ }\bibfield  {title} {\enquote {\bibinfo {title} {{Magnon gap tuning in lithium-doped MnTe}},}\ }\href {https://journals.aps.org/prb/abstract/10.1103/PhysRevB.109.214434} {\bibfield  {journal} {\bibinfo  {journal} {Physical Review B}\ }\textbf {\bibinfo {volume} {109}},\ \bibinfo {pages} {214434} (\bibinfo {year}
  {2024})}\BibitemShut {NoStop}%
\bibitem [{\citenamefont {Mardele}\ \emph {et~al.}(2024)\citenamefont {Mardele}, \citenamefont {Mohelsky}, \citenamefont {Jana}, \citenamefont {Pawbake}, \citenamefont {Dzian}, \citenamefont {Lee}, \citenamefont {Raju}, \citenamefont {Sankar}, \citenamefont {Faugeras}, \citenamefont {Potemski} \emph {et~al.}}]{mardele2024tuning}%
  \BibitemOpen
  \bibfield  {author} {\bibinfo {author} {\bibfnamefont {F.~L.}\ \bibnamefont {Mardele}}, \bibinfo {author} {\bibfnamefont {I.}~\bibnamefont {Mohelsky}}, \bibinfo {author} {\bibfnamefont {D.}~\bibnamefont {Jana}}, \bibinfo {author} {\bibfnamefont {A.}~\bibnamefont {Pawbake}}, \bibinfo {author} {\bibfnamefont {J.}~\bibnamefont {Dzian}}, \bibinfo {author} {\bibfnamefont {W.-L.}\ \bibnamefont {Lee}}, \bibinfo {author} {\bibfnamefont {K.}~\bibnamefont {Raju}}, \bibinfo {author} {\bibfnamefont {R.}~\bibnamefont {Sankar}}, \bibinfo {author} {\bibfnamefont {C.}~\bibnamefont {Faugeras}}, \bibinfo {author} {\bibfnamefont {M.}~\bibnamefont {Potemski}},  \emph {et~al.},\ }\bibfield  {title} {\enquote {\bibinfo {title} {{Tuning THz magnons in a mixed van-der-Waals antiferromagnet}},}\ }\href {https://arxiv.org/abs/2408.12230} {\bibfield  {journal} {\bibinfo  {journal} {arXiv preprint arXiv:2408.12230}\ } (\bibinfo {year} {2024})}\BibitemShut {NoStop}%
\bibitem [{\citenamefont {Vu}\ \emph {et~al.}(2023)\citenamefont {Vu}, \citenamefont {Nelson}, \citenamefont {Wooten}, \citenamefont {Barker}, \citenamefont {Goldberger},\ and\ \citenamefont {Heremans}}]{vu2023magnon}%
  \BibitemOpen
  \bibfield  {author} {\bibinfo {author} {\bibfnamefont {D.~D.}\ \bibnamefont {Vu}}, \bibinfo {author} {\bibfnamefont {R.~A.}\ \bibnamefont {Nelson}}, \bibinfo {author} {\bibfnamefont {B.~L.}\ \bibnamefont {Wooten}}, \bibinfo {author} {\bibfnamefont {J.}~\bibnamefont {Barker}}, \bibinfo {author} {\bibfnamefont {J.~E.}\ \bibnamefont {Goldberger}}, \ and\ \bibinfo {author} {\bibfnamefont {J.~P.}\ \bibnamefont {Heremans}},\ }\bibfield  {title} {\enquote {\bibinfo {title} {{Magnon gap mediated lattice thermal conductivity in MnBi$_2$Te$_4$}},}\ }\href {https://journals.aps.org/prb/abstract/10.1103/PhysRevB.108.144402} {\bibfield  {journal} {\bibinfo  {journal} {Physical Review B}\ }\textbf {\bibinfo {volume} {108}},\ \bibinfo {pages} {144402} (\bibinfo {year} {2023})}\BibitemShut {NoStop}%
\bibitem [{\citenamefont {Tabuchi}\ \emph {et~al.}(2015)\citenamefont {Tabuchi}, \citenamefont {Ishino}, \citenamefont {Noguchi}, \citenamefont {Ishikawa}, \citenamefont {Yamazaki}, \citenamefont {Usami},\ and\ \citenamefont {Nakamura}}]{tabuchi2015coherent}%
  \BibitemOpen
  \bibfield  {author} {\bibinfo {author} {\bibfnamefont {Y.}~\bibnamefont {Tabuchi}}, \bibinfo {author} {\bibfnamefont {S.}~\bibnamefont {Ishino}}, \bibinfo {author} {\bibfnamefont {A.}~\bibnamefont {Noguchi}}, \bibinfo {author} {\bibfnamefont {T.}~\bibnamefont {Ishikawa}}, \bibinfo {author} {\bibfnamefont {R.}~\bibnamefont {Yamazaki}}, \bibinfo {author} {\bibfnamefont {K.}~\bibnamefont {Usami}}, \ and\ \bibinfo {author} {\bibfnamefont {Y.}~\bibnamefont {Nakamura}},\ }\bibfield  {title} {\enquote {\bibinfo {title} {{Coherent coupling between a ferromagnetic magnon and a superconducting qubit}},}\ }\href {https://www.science.org/doi/full/10.1126/science.aaa3693} {\bibfield  {journal} {\bibinfo  {journal} {Science}\ }\textbf {\bibinfo {volume} {349}},\ \bibinfo {pages} {405--408} (\bibinfo {year} {2015})}\BibitemShut {NoStop}%
\bibitem [{\citenamefont {Tabuchi}\ \emph {et~al.}(2016)\citenamefont {Tabuchi}, \citenamefont {Ishino}, \citenamefont {Noguchi}, \citenamefont {Ishikawa}, \citenamefont {Yamazaki}, \citenamefont {Usami},\ and\ \citenamefont {Nakamura}}]{tabuchi2016quantum}%
  \BibitemOpen
  \bibfield  {author} {\bibinfo {author} {\bibfnamefont {Y.}~\bibnamefont {Tabuchi}}, \bibinfo {author} {\bibfnamefont {S.}~\bibnamefont {Ishino}}, \bibinfo {author} {\bibfnamefont {A.}~\bibnamefont {Noguchi}}, \bibinfo {author} {\bibfnamefont {T.}~\bibnamefont {Ishikawa}}, \bibinfo {author} {\bibfnamefont {R.}~\bibnamefont {Yamazaki}}, \bibinfo {author} {\bibfnamefont {K.}~\bibnamefont {Usami}}, \ and\ \bibinfo {author} {\bibfnamefont {Y.}~\bibnamefont {Nakamura}},\ }\bibfield  {title} {\enquote {\bibinfo {title} {{Quantum magnonics: The magnon meets the superconducting qubit}},}\ }\href {https://comptes-rendus.academie-sciences.fr/physique/articles/10.1016/j.crhy.2016.07.009/} {\bibfield  {journal} {\bibinfo  {journal} {Comptes Rendus. Physique}\ }\textbf {\bibinfo {volume} {17}},\ \bibinfo {pages} {729--739} (\bibinfo {year} {2016})}\BibitemShut {NoStop}%
\bibitem [{\citenamefont {Dobrovolskiy}\ \emph {et~al.}(2019)\citenamefont {Dobrovolskiy}, \citenamefont {Sachser}, \citenamefont {Br{\"a}cher}, \citenamefont {B{\"o}ttcher}, \citenamefont {Kruglyak}, \citenamefont {Vovk}, \citenamefont {Shklovskij}, \citenamefont {Huth}, \citenamefont {Hillebrands},\ and\ \citenamefont {Chumak}}]{dobrovolskiy2019magnon}%
  \BibitemOpen
  \bibfield  {author} {\bibinfo {author} {\bibfnamefont {O.~V.}\ \bibnamefont {Dobrovolskiy}}, \bibinfo {author} {\bibfnamefont {R.}~\bibnamefont {Sachser}}, \bibinfo {author} {\bibfnamefont {T.}~\bibnamefont {Br{\"a}cher}}, \bibinfo {author} {\bibfnamefont {T.}~\bibnamefont {B{\"o}ttcher}}, \bibinfo {author} {\bibfnamefont {V.~V.}\ \bibnamefont {Kruglyak}}, \bibinfo {author} {\bibfnamefont {R.~V.}\ \bibnamefont {Vovk}}, \bibinfo {author} {\bibfnamefont {V.~A.}\ \bibnamefont {Shklovskij}}, \bibinfo {author} {\bibfnamefont {M.}~\bibnamefont {Huth}}, \bibinfo {author} {\bibfnamefont {B.}~\bibnamefont {Hillebrands}}, \ and\ \bibinfo {author} {\bibfnamefont {A.~V.}\ \bibnamefont {Chumak}},\ }\bibfield  {title} {\enquote {\bibinfo {title} {{Magnon--fluxon interaction in a ferromagnet/superconductor heterostructure}},}\ }\href {https://www.nature.com/articles/s41567-019-0428-5} {\bibfield  {journal} {\bibinfo  {journal} {Nature Physics}\ }\textbf {\bibinfo {volume} {15}},\ \bibinfo {pages} {477--482} (\bibinfo
  {year} {2019})}\BibitemShut {NoStop}%
\bibitem [{\citenamefont {Niedzielski}\ \emph {et~al.}(2023)\citenamefont {Niedzielski}, \citenamefont {Jia},\ and\ \citenamefont {Berakdar}}]{niedzielski2023magnon}%
  \BibitemOpen
  \bibfield  {author} {\bibinfo {author} {\bibfnamefont {B.}~\bibnamefont {Niedzielski}}, \bibinfo {author} {\bibfnamefont {C.~L.}\ \bibnamefont {Jia}}, \ and\ \bibinfo {author} {\bibfnamefont {J.}~\bibnamefont {Berakdar}},\ }\bibfield  {title} {\enquote {\bibinfo {title} {{Magnon-fluxon interaction in coupled superconductor/ferromagnet hybrid periodic structures}},}\ }\href {https://journals.aps.org/prapplied/abstract/10.1103/PhysRevApplied.19.024073} {\bibfield  {journal} {\bibinfo  {journal} {Physical Review Applied}\ }\textbf {\bibinfo {volume} {19}},\ \bibinfo {pages} {024073} (\bibinfo {year} {2023})}\BibitemShut {NoStop}%
\bibitem [{\citenamefont {Liebhaber}\ \emph {et~al.}(2022)\citenamefont {Liebhaber}, \citenamefont {R{\"u}tten}, \citenamefont {Reecht}, \citenamefont {Steiner}, \citenamefont {Rohlf}, \citenamefont {Rossnagel}, \citenamefont {von Oppen},\ and\ \citenamefont {Franke}}]{liebhaber2022quantum}%
  \BibitemOpen
  \bibfield  {author} {\bibinfo {author} {\bibfnamefont {E.}~\bibnamefont {Liebhaber}}, \bibinfo {author} {\bibfnamefont {L.~M.}\ \bibnamefont {R{\"u}tten}}, \bibinfo {author} {\bibfnamefont {G.}~\bibnamefont {Reecht}}, \bibinfo {author} {\bibfnamefont {J.~F.}\ \bibnamefont {Steiner}}, \bibinfo {author} {\bibfnamefont {S.}~\bibnamefont {Rohlf}}, \bibinfo {author} {\bibfnamefont {K.}~\bibnamefont {Rossnagel}}, \bibinfo {author} {\bibfnamefont {F.}~\bibnamefont {von Oppen}}, \ and\ \bibinfo {author} {\bibfnamefont {K.~J.}\ \bibnamefont {Franke}},\ }\bibfield  {title} {\enquote {\bibinfo {title} {Quantum spins and hybridization in artificially-constructed chains of magnetic adatoms on a superconductor},}\ }\href {https://www.nature.com/articles/s41467-022-29879-0} {\bibfield  {journal} {\bibinfo  {journal} {Nature Communications}\ }\textbf {\bibinfo {volume} {13}},\ \bibinfo {pages} {2160} (\bibinfo {year} {2022})}\BibitemShut {NoStop}%
\bibitem [{\citenamefont {le~Sueur}\ \emph {et~al.}(2008)\citenamefont {le~Sueur}, \citenamefont {Joyez}, \citenamefont {Pothier}, \citenamefont {Urbina},\ and\ \citenamefont {Esteve}}]{lesueur_prl_08}%
  \BibitemOpen
  \bibfield  {author} {\bibinfo {author} {\bibfnamefont {H.}~\bibnamefont {le~Sueur}}, \bibinfo {author} {\bibfnamefont {P.}~\bibnamefont {Joyez}}, \bibinfo {author} {\bibfnamefont {H.}~\bibnamefont {Pothier}}, \bibinfo {author} {\bibfnamefont {C.}~\bibnamefont {Urbina}}, \ and\ \bibinfo {author} {\bibfnamefont {D.}~\bibnamefont {Esteve}},\ }\bibfield  {title} {\enquote {\bibinfo {title} {Phase controlled superconducting proximity effect probed by tunneling spectroscopy},}\ }\href {\doibase 10.1103/PhysRevLett.100.197002} {\bibfield  {journal} {\bibinfo  {journal} {Phys. Rev. Lett.}\ }\textbf {\bibinfo {volume} {100}},\ \bibinfo {pages} {197002} (\bibinfo {year} {2008})}\BibitemShut {NoStop}%
\bibitem [{SM()}]{SM}%
  \BibitemOpen
  \href@noop {} {}\bibinfo {note} {{See Supplementary Material, including references \cite{hodt2024spin,gross1986anomalous,tinkham_book}, at $\langle$URL to be inserted$\rangle$ for the derivation details of the effective spin-spin interaction, the numerical method for determining the ground state configuration, the definition of the origin for the center-of-mass coordinate of the magnetic adatoms, derivation details for the supercurrent-induced magnon gap, the gap equation for the superconducting order parameter, and a suggested setup for tuning the supercurrents.}}\BibitemShut {Stop}%
\bibitem [{\citenamefont {Halterman}\ \emph {et~al.}(2007)\citenamefont {Halterman}, \citenamefont {Barsic},\ and\ \citenamefont {Valls}}]{halterman2007odd}%
  \BibitemOpen
  \bibfield  {author} {\bibinfo {author} {\bibfnamefont {K.}~\bibnamefont {Halterman}}, \bibinfo {author} {\bibfnamefont {P.~H.}\ \bibnamefont {Barsic}}, \ and\ \bibinfo {author} {\bibfnamefont {O.~T.}\ \bibnamefont {Valls}},\ }\bibfield  {title} {\enquote {\bibinfo {title} {Odd triplet pairing in clean superconductor/ferromagnet heterostructures},}\ }\href {https://journals.aps.org/prl/abstract/10.1103/PhysRevLett.99.127002} {\bibfield  {journal} {\bibinfo  {journal} {Physical Review Letters}\ }\textbf {\bibinfo {volume} {99}},\ \bibinfo {pages} {127002} (\bibinfo {year} {2007})}\BibitemShut {NoStop}%
\bibitem [{\citenamefont {Halterman}\ \emph {et~al.}(2008)\citenamefont {Halterman}, \citenamefont {Valls},\ and\ \citenamefont {Barsic}}]{halterman2008induced}%
  \BibitemOpen
  \bibfield  {author} {\bibinfo {author} {\bibfnamefont {K.}~\bibnamefont {Halterman}}, \bibinfo {author} {\bibfnamefont {O.~T.}\ \bibnamefont {Valls}}, \ and\ \bibinfo {author} {\bibfnamefont {P.~H.}\ \bibnamefont {Barsic}},\ }\bibfield  {title} {\enquote {\bibinfo {title} {Induced triplet pairing in clean $s$-wave superconductor/ferromagnet layered structures},}\ }\href {https://journals.aps.org/prb/abstract/10.1103/PhysRevB.77.174511} {\bibfield  {journal} {\bibinfo  {journal} {Physical Review B—Condensed Matter and Materials Physics}\ }\textbf {\bibinfo {volume} {77}},\ \bibinfo {pages} {174511} (\bibinfo {year} {2008})}\BibitemShut {NoStop}%
\bibitem [{\citenamefont {Kalcheim}\ \emph {et~al.}(2015)\citenamefont {Kalcheim}, \citenamefont {Millo}, \citenamefont {Di~Bernardo}, \citenamefont {Pal},\ and\ \citenamefont {Robinson}}]{kalcheim_prb_15}%
  \BibitemOpen
  \bibfield  {author} {\bibinfo {author} {\bibfnamefont {Y.}~\bibnamefont {Kalcheim}}, \bibinfo {author} {\bibfnamefont {O.}~\bibnamefont {Millo}}, \bibinfo {author} {\bibfnamefont {A.}~\bibnamefont {Di~Bernardo}}, \bibinfo {author} {\bibfnamefont {A.}~\bibnamefont {Pal}}, \ and\ \bibinfo {author} {\bibfnamefont {J.~W.~A.}\ \bibnamefont {Robinson}},\ }\bibfield  {title} {\enquote {\bibinfo {title} {Inverse proximity effect at superconductor-ferromagnet interfaces: Evidence for induced triplet pairing in the superconductor},}\ }\href {\doibase 10.1103/PhysRevB.92.060501} {\bibfield  {journal} {\bibinfo  {journal} {Physical Review B}\ }\textbf {\bibinfo {volume} {92}},\ \bibinfo {pages} {060501} (\bibinfo {year} {2015})}\BibitemShut {NoStop}%
\bibitem [{\citenamefont {Grein}\ \emph {et~al.}(2013)\citenamefont {Grein}, \citenamefont {L{\"o}fwander},\ and\ \citenamefont {Eschrig}}]{grein2013inverse}%
  \BibitemOpen
  \bibfield  {author} {\bibinfo {author} {\bibfnamefont {Roland}\ \bibnamefont {Grein}}, \bibinfo {author} {\bibfnamefont {Tomas}\ \bibnamefont {L{\"o}fwander}}, \ and\ \bibinfo {author} {\bibfnamefont {Matthias}\ \bibnamefont {Eschrig}},\ }\bibfield  {title} {\enquote {\bibinfo {title} {Inverse proximity effect and influence of disorder on triplet supercurrents in strongly spin-polarized ferromagnets},}\ }\href {https://journals.aps.org/prb/abstract/10.1103/PhysRevB.88.054502} {\bibfield  {journal} {\bibinfo  {journal} {Physical Review B—Condensed Matter and Materials Physics}\ }\textbf {\bibinfo {volume} {88}},\ \bibinfo {pages} {054502} (\bibinfo {year} {2013})}\BibitemShut {NoStop}%
\bibitem [{\citenamefont {Eschrig}\ and\ \citenamefont {L{\"o}fwander}(2008)}]{eschrig2008triplet}%
  \BibitemOpen
  \bibfield  {author} {\bibinfo {author} {\bibfnamefont {Matthias}\ \bibnamefont {Eschrig}}\ and\ \bibinfo {author} {\bibfnamefont {Tomas}\ \bibnamefont {L{\"o}fwander}},\ }\bibfield  {title} {\enquote {\bibinfo {title} {Triplet supercurrents in clean and disordered half-metallic ferromagnets},}\ }\href {https://www.nature.com/articles/nphys831} {\bibfield  {journal} {\bibinfo  {journal} {Nature Physics}\ }\textbf {\bibinfo {volume} {4}},\ \bibinfo {pages} {138--143} (\bibinfo {year} {2008})}\BibitemShut {NoStop}%
\bibitem [{\citenamefont {Khaire}\ \emph {et~al.}(2010)\citenamefont {Khaire}, \citenamefont {Khasawneh}, \citenamefont {Pratt},\ and\ \citenamefont {Birge}}]{khaire_prl_10}%
  \BibitemOpen
  \bibfield  {author} {\bibinfo {author} {\bibfnamefont {T.~S.}\ \bibnamefont {Khaire}}, \bibinfo {author} {\bibfnamefont {M.~A.}\ \bibnamefont {Khasawneh}}, \bibinfo {author} {\bibfnamefont {W.~P.}\ \bibnamefont {Pratt}}, \ and\ \bibinfo {author} {\bibfnamefont {N.~O.}\ \bibnamefont {Birge}},\ }\bibfield  {title} {\enquote {\bibinfo {title} {{Observation of spin-triplet superconductivity in Co-based Josephson junctions}},}\ }\href {https://journals.aps.org/prl/abstract/10.1103/PhysRevLett.104.137002} {\bibfield  {journal} {\bibinfo  {journal} {Phys. Rev. Lett.}\ }\textbf {\bibinfo {volume} {104}},\ \bibinfo {pages} {137002} (\bibinfo {year} {2010})}\BibitemShut {NoStop}%
\bibitem [{\citenamefont {Robinson}\ \emph {et~al.}(2010)\citenamefont {Robinson}, \citenamefont {Witt},\ and\ \citenamefont {Blamire}}]{robinson_science_10}%
  \BibitemOpen
  \bibfield  {author} {\bibinfo {author} {\bibfnamefont {J.~W.~A.}\ \bibnamefont {Robinson}}, \bibinfo {author} {\bibfnamefont {J.~D.~S.}\ \bibnamefont {Witt}}, \ and\ \bibinfo {author} {\bibfnamefont {M.~G.}\ \bibnamefont {Blamire}},\ }\bibfield  {title} {\enquote {\bibinfo {title} {Controlled injection of spin-triplet supercurrents into a strong ferromagnet},}\ }\href {\doibase 10.1126/science.1189246} {\bibfield  {journal} {\bibinfo  {journal} {Science}\ }\textbf {\bibinfo {volume} {329}},\ \bibinfo {pages} {59--61} (\bibinfo {year} {2010})}\BibitemShut {NoStop}%
\bibitem [{\citenamefont {Reeg}\ and\ \citenamefont {Maslov}(2015)}]{reeg2015proximity}%
  \BibitemOpen
  \bibfield  {author} {\bibinfo {author} {\bibfnamefont {C.~R.}\ \bibnamefont {Reeg}}\ and\ \bibinfo {author} {\bibfnamefont {D.~L.}\ \bibnamefont {Maslov}},\ }\bibfield  {title} {\enquote {\bibinfo {title} {Proximity-induced triplet superconductivity in {Rashba} materials},}\ }\href {https://journals.aps.org/prb/abstract/10.1103/PhysRevB.92.134512} {\bibfield  {journal} {\bibinfo  {journal} {Physical Review B}\ }\textbf {\bibinfo {volume} {92}},\ \bibinfo {pages} {134512} (\bibinfo {year} {2015})}\BibitemShut {NoStop}%
\bibitem [{\citenamefont {Ying}\ \emph {et~al.}(2017)\citenamefont {Ying}, \citenamefont {Cuoco}, \citenamefont {Ortix},\ and\ \citenamefont {Gentile}}]{ying2017tuning}%
  \BibitemOpen
  \bibfield  {author} {\bibinfo {author} {\bibfnamefont {Zu-Jian}\ \bibnamefont {Ying}}, \bibinfo {author} {\bibfnamefont {Mario}\ \bibnamefont {Cuoco}}, \bibinfo {author} {\bibfnamefont {Carmine}\ \bibnamefont {Ortix}}, \ and\ \bibinfo {author} {\bibfnamefont {Paola}\ \bibnamefont {Gentile}},\ }\bibfield  {title} {\enquote {\bibinfo {title} {Tuning pairing amplitude and spin-triplet texture by curving superconducting nanostructures},}\ }\href {https://journals.aps.org/prb/abstract/10.1103/PhysRevB.96.100506} {\bibfield  {journal} {\bibinfo  {journal} {Physical Review B}\ }\textbf {\bibinfo {volume} {96}},\ \bibinfo {pages} {100506} (\bibinfo {year} {2017})}\BibitemShut {NoStop}%
\bibitem [{\citenamefont {Gassner}\ \emph {et~al.}(2024)\citenamefont {Gassner}, \citenamefont {Weber},\ and\ \citenamefont {Claassen}}]{gassner2024light}%
  \BibitemOpen
  \bibfield  {author} {\bibinfo {author} {\bibfnamefont {S.}~\bibnamefont {Gassner}}, \bibinfo {author} {\bibfnamefont {C.~S.}\ \bibnamefont {Weber}}, \ and\ \bibinfo {author} {\bibfnamefont {M.}~\bibnamefont {Claassen}},\ }\bibfield  {title} {\enquote {\bibinfo {title} {Light-induced switching between singlet and triplet superconducting states},}\ }\href {https://www.nature.com/articles/s41467-024-45949-x} {\bibfield  {journal} {\bibinfo  {journal} {Nature Communications}\ }\textbf {\bibinfo {volume} {15}},\ \bibinfo {pages} {1776} (\bibinfo {year} {2024})}\BibitemShut {NoStop}%
\bibitem [{\citenamefont {Takashima}\ \emph {et~al.}(2017)\citenamefont {Takashima}, \citenamefont {Fujimoto},\ and\ \citenamefont {Yokoyama}}]{Takashima2017:PRB}%
  \BibitemOpen
  \bibfield  {author} {\bibinfo {author} {\bibfnamefont {R.}~\bibnamefont {Takashima}}, \bibinfo {author} {\bibfnamefont {S.}~\bibnamefont {Fujimoto}}, \ and\ \bibinfo {author} {\bibfnamefont {T.}~\bibnamefont {Yokoyama}},\ }\bibfield  {title} {\enquote {\bibinfo {title} {Adiabatic and nonadiabatic spin torques induced by a spin-triplet supercurrent},}\ }\href {https://journals.aps.org/prb/abstract/10.1103/PhysRevB.96.121203} {\bibfield  {journal} {\bibinfo  {journal} {Physical Review B}\ }\textbf {\bibinfo {volume} {96}},\ \bibinfo {pages} {121203(R)} (\bibinfo {year} {2017})}\BibitemShut {NoStop}%
\bibitem [{\citenamefont {Balatsky}\ \emph {et~al.}(2006)\citenamefont {Balatsky}, \citenamefont {Vekhter},\ and\ \citenamefont {Zhu}}]{balatsky_rmp_06}%
  \BibitemOpen
  \bibfield  {author} {\bibinfo {author} {\bibfnamefont {A.~V.}\ \bibnamefont {Balatsky}}, \bibinfo {author} {\bibfnamefont {I.}~\bibnamefont {Vekhter}}, \ and\ \bibinfo {author} {\bibfnamefont {Jian-Xin}\ \bibnamefont {Zhu}},\ }\bibfield  {title} {\enquote {\bibinfo {title} {Impurity-induced states in conventional and unconventional superconductors},}\ }\href {\doibase 10.1103/RevModPhys.78.373} {\bibfield  {journal} {\bibinfo  {journal} {Rev. Mod. Phys.}\ }\textbf {\bibinfo {volume} {78}},\ \bibinfo {pages} {373--433} (\bibinfo {year} {2006})}\BibitemShut {NoStop}%
\bibitem [{\citenamefont {Deb}\ \emph {et~al.}(2021)\citenamefont {Deb}, \citenamefont {Hoffman}, \citenamefont {Loss},\ and\ \citenamefont {Klinovaja}}]{deb_prb_21}%
  \BibitemOpen
  \bibfield  {author} {\bibinfo {author} {\bibfnamefont {Oindrila}\ \bibnamefont {Deb}}, \bibinfo {author} {\bibfnamefont {Silas}\ \bibnamefont {Hoffman}}, \bibinfo {author} {\bibfnamefont {Daniel}\ \bibnamefont {Loss}}, \ and\ \bibinfo {author} {\bibfnamefont {Jelena}\ \bibnamefont {Klinovaja}},\ }\bibfield  {title} {\enquote {\bibinfo {title} {{Yu-Shiba-Rusinov states and ordering of magnetic impurities near the boundary of a superconducting nanowire}},}\ }\href {\doibase 10.1103/PhysRevB.103.165403} {\bibfield  {journal} {\bibinfo  {journal} {Phys. Rev. B}\ }\textbf {\bibinfo {volume} {103}},\ \bibinfo {pages} {165403} (\bibinfo {year} {2021})}\BibitemShut {NoStop}%
\bibitem [{\citenamefont {Meservey}\ and\ \citenamefont {Tedrow}(1994)}]{meservey_physrep_94}%
  \BibitemOpen
  \bibfield  {author} {\bibinfo {author} {\bibfnamefont {R.}~\bibnamefont {Meservey}}\ and\ \bibinfo {author} {\bibfnamefont {P.~M.}\ \bibnamefont {Tedrow}},\ }\bibfield  {title} {\enquote {\bibinfo {title} {Spin-polarized electron tunneling},}\ }\href {https://www.sciencedirect.com/science/article/abs/pii/0370157394901058} {\bibfield  {journal} {\bibinfo  {journal} {Phys. Rep.}\ }\textbf {\bibinfo {volume} {238}},\ \bibinfo {pages} {173--243} (\bibinfo {year} {1994})}\BibitemShut {NoStop}%
\bibitem [{\citenamefont {Lo~Conte}\ \emph {et~al.}(2024)\citenamefont {Lo~Conte}, \citenamefont {Wiebe}, \citenamefont {Rachel}, \citenamefont {Morr},\ and\ \citenamefont {Wiesendanger}}]{conte_arxiv_24}%
  \BibitemOpen
  \bibfield  {author} {\bibinfo {author} {\bibfnamefont {R.}~\bibnamefont {Lo~Conte}}, \bibinfo {author} {\bibfnamefont {J.}~\bibnamefont {Wiebe}}, \bibinfo {author} {\bibfnamefont {S.}~\bibnamefont {Rachel}}, \bibinfo {author} {\bibfnamefont {D.~K.}\ \bibnamefont {Morr}}, \ and\ \bibinfo {author} {\bibfnamefont {R.}~\bibnamefont {Wiesendanger}},\ }\bibfield  {title} {\enquote {\bibinfo {title} {{Magnet-superconductor hybrid quantum systems: a materials platform for topological superconductivity}},}\ }\href {\doibase 10.48550/arXiv.2410.2017} {\bibfield  {journal} {\bibinfo  {journal} {arXiv}\ } (\bibinfo {year} {2024}),\ 10.48550/arXiv.2410.2017},\ \Eprint {http://arxiv.org/abs/cond-mat/2410.20177} {cond-mat/2410.20177} \BibitemShut {NoStop}%
\bibitem [{\citenamefont {Christensen}\ \emph {et~al.}(2016)\citenamefont {Christensen}, \citenamefont {Schecter}, \citenamefont {Flensberg}, \citenamefont {Andersen},\ and\ \citenamefont {Paaske}}]{christensen2016spiral}%
  \BibitemOpen
  \bibfield  {author} {\bibinfo {author} {\bibfnamefont {Morten~H}\ \bibnamefont {Christensen}}, \bibinfo {author} {\bibfnamefont {Michael}\ \bibnamefont {Schecter}}, \bibinfo {author} {\bibfnamefont {Karsten}\ \bibnamefont {Flensberg}}, \bibinfo {author} {\bibfnamefont {Brian~M}\ \bibnamefont {Andersen}}, \ and\ \bibinfo {author} {\bibfnamefont {Jens}\ \bibnamefont {Paaske}},\ }\bibfield  {title} {\enquote {\bibinfo {title} {Spiral magnetic order and topological superconductivity in a chain of magnetic adatoms on a two-dimensional superconductor},}\ }\href {https://journals.aps.org/prb/abstract/10.1103/PhysRevB.94.144509} {\bibfield  {journal} {\bibinfo  {journal} {Physical Review B}\ }\textbf {\bibinfo {volume} {94}},\ \bibinfo {pages} {144509} (\bibinfo {year} {2016})}\BibitemShut {NoStop}%
\bibitem [{\citenamefont {Pientka}\ \emph {et~al.}(2013)\citenamefont {Pientka}, \citenamefont {Glazman},\ and\ \citenamefont {Von~Oppen}}]{pientka2013topological}%
  \BibitemOpen
  \bibfield  {author} {\bibinfo {author} {\bibfnamefont {Falko}\ \bibnamefont {Pientka}}, \bibinfo {author} {\bibfnamefont {Leonid~I}\ \bibnamefont {Glazman}}, \ and\ \bibinfo {author} {\bibfnamefont {Felix}\ \bibnamefont {Von~Oppen}},\ }\bibfield  {title} {\enquote {\bibinfo {title} {Topological superconducting phase in helical shiba chains},}\ }\href {https://journals.aps.org/prb/abstract/10.1103/PhysRevB.88.155420} {\bibfield  {journal} {\bibinfo  {journal} {Physical Review B—Condensed Matter and Materials Physics}\ }\textbf {\bibinfo {volume} {88}},\ \bibinfo {pages} {155420} (\bibinfo {year} {2013})}\BibitemShut {NoStop}%
\bibitem [{\citenamefont {Awoga}\ \emph {et~al.}(2024)\citenamefont {Awoga}, \citenamefont {Ioannidis}, \citenamefont {Mishra}, \citenamefont {Leijnse}, \citenamefont {Trif},\ and\ \citenamefont {Posske}}]{awoga2024controlling}%
  \BibitemOpen
  \bibfield  {author} {\bibinfo {author} {\bibfnamefont {Oladunjoye~A}\ \bibnamefont {Awoga}}, \bibinfo {author} {\bibfnamefont {Ioannis}\ \bibnamefont {Ioannidis}}, \bibinfo {author} {\bibfnamefont {Archana}\ \bibnamefont {Mishra}}, \bibinfo {author} {\bibfnamefont {Martin}\ \bibnamefont {Leijnse}}, \bibinfo {author} {\bibfnamefont {Mircea}\ \bibnamefont {Trif}}, \ and\ \bibinfo {author} {\bibfnamefont {Thore}\ \bibnamefont {Posske}},\ }\bibfield  {title} {\enquote {\bibinfo {title} {{Controlling Majorana hybridization in magnetic chain-superconductor systems}},}\ }\href {https://journals.aps.org/prresearch/abstract/10.1103/PhysRevResearch.6.033154} {\bibfield  {journal} {\bibinfo  {journal} {Physical Review Research}\ }\textbf {\bibinfo {volume} {6}},\ \bibinfo {pages} {033154} (\bibinfo {year} {2024})}\BibitemShut {NoStop}%
\bibitem [{\citenamefont {Li}\ \emph {et~al.}(2016)\citenamefont {Li}, \citenamefont {Neupert}, \citenamefont {Wang}, \citenamefont {MacDonald}, \citenamefont {Yazdani},\ and\ \citenamefont {Bernevig}}]{li2016two}%
  \BibitemOpen
  \bibfield  {author} {\bibinfo {author} {\bibfnamefont {Jian}\ \bibnamefont {Li}}, \bibinfo {author} {\bibfnamefont {Titus}\ \bibnamefont {Neupert}}, \bibinfo {author} {\bibfnamefont {Zhijun}\ \bibnamefont {Wang}}, \bibinfo {author} {\bibfnamefont {AH}~\bibnamefont {MacDonald}}, \bibinfo {author} {\bibfnamefont {Ali}\ \bibnamefont {Yazdani}}, \ and\ \bibinfo {author} {\bibfnamefont {B~Andrei}\ \bibnamefont {Bernevig}},\ }\bibfield  {title} {\enquote {\bibinfo {title} {{Two-dimensional chiral topological superconductivity in Shiba lattices}},}\ }\href {https://www.nature.com/articles/ncomms12297} {\bibfield  {journal} {\bibinfo  {journal} {Nature Communications}\ }\textbf {\bibinfo {volume} {7}},\ \bibinfo {pages} {12297} (\bibinfo {year} {2016})}\BibitemShut {NoStop}%
\bibitem [{\citenamefont {Bergeret}\ \emph {et~al.}(2018)\citenamefont {Bergeret}, \citenamefont {Silaev}, \citenamefont {Virtanen},\ and\ \citenamefont {Heikkil\"a}}]{bergeret_rmp_18}%
  \BibitemOpen
  \bibfield  {author} {\bibinfo {author} {\bibfnamefont {F.~S.}\ \bibnamefont {Bergeret}}, \bibinfo {author} {\bibfnamefont {M.}~\bibnamefont {Silaev}}, \bibinfo {author} {\bibfnamefont {P.}~\bibnamefont {Virtanen}}, \ and\ \bibinfo {author} {\bibfnamefont {T.~T.}\ \bibnamefont {Heikkil\"a}},\ }\bibfield  {title} {\enquote {\bibinfo {title} {Colloquium: Nonequilibrium effects in superconductors with a spin-splitting field},}\ }\href {\doibase 10.1103/RevModPhys.90.041001} {\bibfield  {journal} {\bibinfo  {journal} {Rev. Mod. Phys.}\ }\textbf {\bibinfo {volume} {90}},\ \bibinfo {pages} {041001} (\bibinfo {year} {2018})}\BibitemShut {NoStop}%
\bibitem [{\citenamefont {Clogston}(1962)}]{clogston_prl_62}%
  \BibitemOpen
  \bibfield  {author} {\bibinfo {author} {\bibfnamefont {A.~M.}\ \bibnamefont {Clogston}},\ }\bibfield  {title} {\enquote {\bibinfo {title} {Upper {{Limit}} for the {{Critical Field}} in {{Hard Superconductors}}},}\ }\href {\doibase 10.1103/PhysRevLett.9.266} {\bibfield  {journal} {\bibinfo  {journal} {Phys. Rev. Lett.}\ }\textbf {\bibinfo {volume} {9}},\ \bibinfo {pages} {266--267} (\bibinfo {year} {1962})}\BibitemShut {NoStop}%
\bibitem [{\citenamefont {Chandrasekhar}(1962)}]{chandrasekhar_apl_62}%
  \BibitemOpen
  \bibfield  {author} {\bibinfo {author} {\bibfnamefont {B.~S.}\ \bibnamefont {Chandrasekhar}},\ }\href@noop {} {\bibfield  {journal} {\bibinfo  {journal} {Appl. Phys. Lett.}\ }\textbf {\bibinfo {volume} {1}},\ \bibinfo {pages} {7} (\bibinfo {year} {1962})}\BibitemShut {NoStop}%
\bibitem [{\citenamefont {Johnsen}\ \emph {et~al.}(2021)\citenamefont {Johnsen}, \citenamefont {Simensen}, \citenamefont {Brataas},\ and\ \citenamefont {Linder}}]{johnsen2021magnon}%
  \BibitemOpen
  \bibfield  {author} {\bibinfo {author} {\bibfnamefont {L.~G.}\ \bibnamefont {Johnsen}}, \bibinfo {author} {\bibfnamefont {H.~T.}\ \bibnamefont {Simensen}}, \bibinfo {author} {\bibfnamefont {A.}~\bibnamefont {Brataas}}, \ and\ \bibinfo {author} {\bibfnamefont {J.}~\bibnamefont {Linder}},\ }\bibfield  {title} {\enquote {\bibinfo {title} {Magnon spin current induced by triplet {Cooper} pair supercurrents},}\ }\href {https://journals.aps.org/prl/abstract/10.1103/PhysRevLett.127.207001} {\bibfield  {journal} {\bibinfo  {journal} {Physical Review Letters}\ }\textbf {\bibinfo {volume} {127}},\ \bibinfo {pages} {207001} (\bibinfo {year} {2021})}\BibitemShut {NoStop}%
\bibitem [{\citenamefont {Ianovskaia}\ \emph {et~al.}(2023)\citenamefont {Ianovskaia}, \citenamefont {Bobkov},\ and\ \citenamefont {Bobkova}}]{ianovskaia_prb_23}%
  \BibitemOpen
  \bibfield  {author} {\bibinfo {author} {\bibfnamefont {A.~S.}\ \bibnamefont {Ianovskaia}}, \bibinfo {author} {\bibfnamefont {A.~M.}\ \bibnamefont {Bobkov}}, \ and\ \bibinfo {author} {\bibfnamefont {I.~V.}\ \bibnamefont {Bobkova}},\ }\bibfield  {title} {\enquote {\bibinfo {title} {Magnon influence on the superconducting density of states in superconductor--ferromagnetic-insulator bilayers},}\ }\href {\doibase 10.1103/PhysRevB.108.214501} {\bibfield  {journal} {\bibinfo  {journal} {Phys. Rev. B}\ }\textbf {\bibinfo {volume} {108}},\ \bibinfo {pages} {214501} (\bibinfo {year} {2023})}\BibitemShut {NoStop}%
\bibitem [{\citenamefont {Rezende}\ \emph {et~al.}(2019)\citenamefont {Rezende}, \citenamefont {Azevedo},\ and\ \citenamefont {Rodríguez-Suárez}}]{rezende_jap_19}%
  \BibitemOpen
  \bibfield  {author} {\bibinfo {author} {\bibfnamefont {S.~M.}\ \bibnamefont {Rezende}}, \bibinfo {author} {\bibfnamefont {A.}~\bibnamefont {Azevedo}}, \ and\ \bibinfo {author} {\bibfnamefont {R.~L.}\ \bibnamefont {Rodríguez-Suárez}},\ }\bibfield  {title} {\enquote {\bibinfo {title} {Introduction to antiferromagnetic magnons},}\ }\href {\doibase 10.1063/1.5109132} {\bibfield  {journal} {\bibinfo  {journal} {Journal of Applied Physics}\ }\textbf {\bibinfo {volume} {126}},\ \bibinfo {pages} {151101} (\bibinfo {year} {2019})}\BibitemShut {NoStop}%
\bibitem [{\citenamefont {Hodt}\ and\ \citenamefont {Linder}(2024)}]{hodt2024spin}%
  \BibitemOpen
  \bibfield  {author} {\bibinfo {author} {\bibfnamefont {E.~W.}\ \bibnamefont {Hodt}}\ and\ \bibinfo {author} {\bibfnamefont {J.}~\bibnamefont {Linder}},\ }\bibfield  {title} {\enquote {\bibinfo {title} {Spin pumping in an altermagnet/normal-metal bilayer},}\ }\href {https://journals.aps.org/prb/abstract/10.1103/PhysRevB.109.174438} {\bibfield  {journal} {\bibinfo  {journal} {Physical Review B}\ }\textbf {\bibinfo {volume} {109}},\ \bibinfo {pages} {174438} (\bibinfo {year} {2024})}\BibitemShut {NoStop}%
\bibitem [{\citenamefont {Gross}\ \emph {et~al.}(1986)\citenamefont {Gross}, \citenamefont {Chandrasekhar}, \citenamefont {Einzel}, \citenamefont {Andres}, \citenamefont {Hirschfeld}, \citenamefont {Ott}, \citenamefont {Beuers}, \citenamefont {Fisk},\ and\ \citenamefont {Smith}}]{gross1986anomalous}%
  \BibitemOpen
  \bibfield  {author} {\bibinfo {author} {\bibfnamefont {F.}~\bibnamefont {Gross}}, \bibinfo {author} {\bibfnamefont {B.~S.}\ \bibnamefont {Chandrasekhar}}, \bibinfo {author} {\bibfnamefont {D.}~\bibnamefont {Einzel}}, \bibinfo {author} {\bibfnamefont {K.}~\bibnamefont {Andres}}, \bibinfo {author} {\bibfnamefont {P.~J.}\ \bibnamefont {Hirschfeld}}, \bibinfo {author} {\bibfnamefont {H.~R.}\ \bibnamefont {Ott}}, \bibinfo {author} {\bibfnamefont {J.}~\bibnamefont {Beuers}}, \bibinfo {author} {\bibfnamefont {Z.}~\bibnamefont {Fisk}}, \ and\ \bibinfo {author} {\bibfnamefont {J.~L.}\ \bibnamefont {Smith}},\ }\bibfield  {title} {\enquote {\bibinfo {title} {{Anomalous temperature dependence of the magnetic field penetration depth in superconducting UBe$_{13}$}},}\ }\href {https://link.springer.com/article/10.1007/BF01303700} {\bibfield  {journal} {\bibinfo  {journal} {Zeitschrift für Physik B Condensed Matter}\ }\textbf {\bibinfo {volume} {64}},\ \bibinfo {pages} {175--188} (\bibinfo {year} {1986})}\BibitemShut
  {NoStop}%
\bibitem [{\citenamefont {Tinkham}(2004)}]{tinkham_book}%
  \BibitemOpen
  \bibfield  {author} {\bibinfo {author} {\bibfnamefont {M.}~\bibnamefont {Tinkham}},\ }\href@noop {} {\emph {\bibinfo {title} {{Introduction to Superconductivity}}}}\ (\bibinfo  {publisher} {Dover Books on Physics Series, Dover, New York},\ \bibinfo {year} {2004})\BibitemShut {NoStop}%
\bibitem [{\citenamefont {Sigrist}(2005)}]{sigrist2005introduction}%
  \BibitemOpen
  \bibfield  {author} {\bibinfo {author} {\bibfnamefont {M.}~\bibnamefont {Sigrist}},\ }\bibfield  {title} {\enquote {\bibinfo {title} {Introduction to unconventional superconductivity},}\ }\bibfield  {booktitle} {\emph {\bibinfo {booktitle} {AIP Conference Proceedings}},\ }\href {https://pubs.aip.org/aip/acp/article-abstract/1162/1/55/819220/Introduction-to-unconventional-superconductivity} {\ \textbf {\bibinfo {volume} {789}},\ \bibinfo {pages} {165--243} (\bibinfo {year} {2005})}\BibitemShut {NoStop}%
\end{thebibliography}%

\clearpage

\makeatletter 
\renewcommand{\thefigure}{S\@arabic\c@figure}
\makeatother
\setcounter{figure}{0}    

\renewcommand{\theequation}{S\arabic{equation}}
\setcounter{equation}{0}

\onecolumngrid
\begin{center}
    {\large \textbf{Supplemental material}} \\[.2cm]
    Johanne Bratland Tjernshaugen$^1$, Martin Tang Bruland$^1$, and Jacob Linder$^1$ \\[.1cm]
    $^1$\textit{Center for Quantum Spintronics, Department of Physics, Norwegian \\ University of Science and Technology, NO-7491 Trondheim, Norway}
    
\end{center}

\section{Derivation of the effective spin-spin interaction}

\begingroup

\setlength{\tabcolsep}{6pt} 
\renewcommand{\arraystretch}{2} 

\subsection{Diagonalizing the Hamiltonian}
The mean-field Hamiltonian for a supercurrent-carrying triplet superconductor with equal-spin Cooper pairs is
\begin{equation}
\begin{split}
    H_\text{SC} = \sum_{\veck \in \text{BZ}, \sigma} \epsilon_{\veck} c_{\veck, \sigma}^{\dagger} c_{\veck, \sigma} + \frac{1}{2}\sum_{\veck\in \text{BZ}, \sigma} \left( \Delta_{\veck}^{\sigma}\right)^{\dagger}c_{-\veck+\vecQ^{\sigma},\sigma}c_{\veck+\vecQ^{\sigma} }\text{+h.c.} 
\end{split}
\end{equation}
Here, $k$ runs through the crystallographic Brillouin zone (BZ). From this point, we implicitly assume that all $k$- and $q$-sums run over the crystallographic Brillouin zone except if stated otherwise and we write $\Sigma_k$ instead of $\Sigma_{k\in\text{BZ}}$.
By introducing the basis $\phi_{\veck,\sigma}=\begin{pmatrix}c_{\veck+\vecQ^{\sigma}, \sigma} & c^{\dagger}_{-\veck+\vecQ^{\sigma},\sigma}\end{pmatrix}^T$, the Hamiltonian becomes
\begin{equation}
    H_{\text{SC}} = \frac{1}{2}\sum_{\veck, \sigma}\phi^\dagger_{\veck, \sigma}\begin{pmatrix}
        \epsilon_{\veck+\vecQ^{\sigma}} & \Delta_{\veck}^{\sigma} \\ (\Delta_{\veck}^{\sigma})^* & -\epsilon_{-\veck+\vecQ^{\sigma}}
    \end{pmatrix}\phi_{\veck, \sigma}.
\end{equation}
Diagonalization is achieved through the Bogoliubov transformation 
\begin{equation}\label{eq:bog transf equal spin pairing}
    \begin{pmatrix}\gamma_{\veck+\vecQ^{\sigma}, \sigma} \\ \gamma^{\dagger}_{-\veck+\vecQ^{\sigma},\sigma}\end{pmatrix}=\begin{pmatrix}
        u_{\veck, \sigma} & v_{\veck, \sigma} \\ -v^*_{\veck, \sigma} & u^*_{\veck, \sigma}
    \end{pmatrix}\begin{pmatrix}c_{\veck+\vecQ^{\sigma}, \sigma} \\ c^{\dagger}_{-\veck+\vecQ^{\sigma},\sigma}\end{pmatrix},
\end{equation}
where $|u_{\veck,\sigma}|^2+|v_{\veck,\sigma}|^2=1$ for the $\gamma$-operators to be fermionic, and $u_{\veck,\sigma}=u_{-\veck,\sigma}$ and $v_{\veck,\sigma}=-v_{-\veck,\sigma}$ for the transformation to be consistent.  The diagonalized Hamiltonian is $H_\text{SC}=\sum_{k,\sigma} E_{k,\sigma} \gamma_{k,\sigma}^{\dagger}\gamma_{k,\sigma}$. This gives the coherence factors
\begin{equation}\label{eq:coherence factors equal spin pairing}
    u_{\veck,\sigma}=\sqrt{\frac{1}{2}+\frac{1}{2}\frac{\epsilon_{\veck+\vecQ^{\sigma}}+\epsilon_{-\veck+\vecQ^{\sigma}}}{\sqrt{(\epsilon_{\veck+\vecQ^{\sigma}}+\epsilon_{-\veck+\vecQ^{\sigma}})^2+4|\Delta_{\veck}^{\sigma}|^2}}},
    \quad
    v_{\veck,\sigma}=\frac{\Delta_{\veck}^{\sigma}}{|\Delta_{\veck}^{\sigma}|}\sqrt{\frac{1}{2}-\frac{1}{2}\frac{\epsilon_{\veck+\vecQ^{\sigma}}+\epsilon_{-\veck+\vecQ^{\sigma}}}{\sqrt{(\epsilon_{\veck+\vecQ^{\sigma}}+\epsilon_{-\veck+\vecQ^{\sigma}})^2+4|\Delta_{\veck}^{\sigma}|^2}}},
\end{equation}
and the eigenvalues
\begin{equation}\label{eq:eigenvalues equal spin pairing}
\begin{split}
    E_{\veck, \sigma}&=\frac{\epsilon_{\veck}-\epsilon_{-\veck+2\vecQ^{\sigma}}}{2} + \sqrt{\bigg(\frac{\epsilon_{\veck}+\epsilon_{-\veck+2\vecQ^{\sigma}}}{2}\bigg)^2+ |\Delta^{\sigma}_{\veck-\vecQ^{\sigma}}|^2}.
\end{split}
\end{equation}
For later convenience, we define $\Tilde{E}_{\veck, \sigma} = E_{\veck+\vecQ^{\sigma},\sigma}$.

\newpage
\subsection{Effective spin-spin interaction}
The coupling between the itinerant electrons in the superconductor and the classical spins is given by
\begin{equation}
\begin{split}
   H_c=\sum_{\sigma, \sigma'}\sum_{\veck, \vecq } T_{q,\sigma,\sigma'} c^\dagger_{\veck+\vecq,\sigma}c_{\veck, \sigma'}\quad \text{where } \quad T_{q,\sigma\sigma'}=\frac{\Lambda}{{2}N}\sum_{i}\vecS_i\cdot\boldsymbol{\sigma}_{\sigma, \sigma'}\text{e}^{-\text{i}\vecq\cdot{r}_i}.
\end{split}
\end{equation}
The goal is to perform a Schrieffer-Wolff transformation to get an effective spin model. The full Hamiltonian is $H=H_c + H_\text{SC}$, where we treat $H_c$ as a perturbation. Note that diagonal elements of the perturbation $H_c$ should be absorbed into $H_\text{SC}$ in order to allow for first-order contributions in the transformed Hamiltonian. In other words, if $\langle H_c\rangle_\gamma \neq 0$, these terms should be included in $H_\text{SC}$. $\langle H_c\rangle_\gamma$ is calculated in Eq. \eqref{eq:expectation value of Hc}, and in the presence of a pure charge or spin supercurrent, this term is zero. For a spin-polarized charge supercurrent, the first-order term, which is effectively a magnetic field, will dominate the spin Hamiltonian. From here on, we assume that the interaction between the magnetic adatoms is induced by a pure charge or spin supercurrent, and we calculate the spin-spin interaction to second order in $\Lambda$.
The Hamiltonian is canonically transformed to  $H'=\text{e}^{\text{i}S}H\text{e}^{-\text{i}S}$, where $S$ is a hermitian operator known as a generator. This transformed Hamiltonian has the same spectrum as the original Hamiltonian. If the generator is chosen such that $ [S, H_\text{SC}]=\text{i}H_c$, the transformed Hamiltonian is $ H' = H_\text{SC} + \frac{\text{i}}{2}[S, H_c]$ up to second order in $\Lambda$. The effective spin Hamiltonian $\mathcal{H}_\text{eff}$ is found by integrating out the $\gamma$--fermions that diagonalize $H_\text{SC}$: 
\begin{equation}
\mathcal{H}_\text{eff} = \langle \frac{\text{i}}{2}[S, H_c] \rangle_{\gamma}.
\end{equation}

The first step is to find a generator $S$ such that $ [S, H_\text{SC}]=\text{i}H_c$ is satisfied. The coupling term is rewritten in terms of the $\gamma$ operators,
\begin{equation}
\begin{split}
    H_c = \sum_{\vecq,\sigma,\sigma'} T_{\vecq,\sigma,\sigma'} \sum_{\veck}\bigg(u_{\veck+\vecq-\vecQ^{\sigma},\sigma}u^*_{\veck-\vecQ^{\sigma'}, \sigma'}\gamma^{\dagger}_{\veck+\vecq,\sigma}\gamma_{\veck, \sigma'} -u_{\veck+\vecq-\vecQ^{\sigma},\sigma} v_{\veck-\vecQ^{\sigma'},\sigma'}  \gamma^{\dagger}_{\veck+\vecq,\sigma}\gamma^{\dagger}_{-\veck+2\vecQ^{\sigma'},\sigma'} \\ -v^*_{\veck+\vecq-\vecQ^{\sigma},\sigma}u^*_{\veck-\vecQ^{\sigma'},\sigma'}\gamma_{-\veck-\vecq+2\vecQ^{\sigma},\sigma}\gamma_{\veck, \sigma'} + v^*_{\veck+\vecq-\vecQ^{\sigma},\sigma}v_{\veck-\vecQ^{\sigma'},\sigma'}\gamma_{-\veck-\vecq+2\vecQ^{\sigma},\sigma}\gamma^{\dagger}_{-\veck+2\vecQ^{\sigma'}, \sigma'} \bigg),
\end{split}
\end{equation}
and the ansatz generator $S$ has the same form but with different prefactors,
\begin{equation}
\begin{split}
    S = &\sum_{\veck, \vecq,\sigma,\sigma'} \bigg(A_{\veck, \vecq,\sigma,\sigma'}\gamma^{\dagger}_{\veck+\vecq,\sigma}\gamma_{\veck, \sigma'}  +B_{\veck, \vecq,\sigma,\sigma'}  \gamma^{\dagger}_{\veck+\vecq,\sigma}\gamma^{\dagger}_{-\veck+2\vecQ^{\sigma'},\sigma'} \\ &+C_{\veck, \vecq,\sigma,\sigma'}\gamma_{-\veck-\vecq+2\vecQ^{\sigma},\sigma}\gamma_{\veck, \sigma'} + D_{\veck, \vecq,\sigma,\sigma'}\gamma_{-\veck-\vecq+2\vecQ^{\sigma},\sigma}\gamma^{\dagger}_{-\veck+2\vecQ^{\sigma'}, \sigma'} \bigg).
\end{split}
\end{equation}
The prefactors are determined by comparing the left-hand side of the equation $[S, H_\text{SC}]=\text{i}H_c$ to the right-hand side. This gives expressions of the kind 
\begin{equation}
\begin{split}
    A_{\veck, \vecq,\sigma,\sigma'}(E_{\veck, \sigma'} - E_{\veck+\vecq, \sigma}) = \text{i}T_{\vecq,\sigma,\sigma'}u_{\veck+\vecq-\vecQ^{\sigma},\sigma}u^*_{\veck-\vecQ^{\sigma'},\sigma'}.
\end{split}
\end{equation}
The right-hand side of this equation is in general nonzero, even when $E_{\veck, \sigma'} = E_{\veck+\vecq, \sigma}$. This is resolved by letting $ A_{\veck, \vecq,\sigma,\sigma'}$ diverge such that the equality is satisfied. In the final expression for the Hamiltonian, $ A_{\veck, \vecq,\sigma,\sigma'}$ always shows up in combination with Fermi-Dirac distributions. This gives a derivative of the Fermi-Dirac distribution as $E_{\veck, \sigma'} \rightarrow E_{\veck+\vecq, \sigma}$:
\begin{equation}
     A_{\veck, \vecq,\sigma,\sigma'}[n(E_{\veck, \sigma'}) - n(E_{\veck+\vecq, \sigma})] = \text{i}T_{\vecq,\sigma,\sigma'}u_{\veck+\vecq-\vecQ^{\sigma},\sigma}u^*_{\veck-\vecQ^{\sigma'},\sigma'} n'(E_{\veck, \sigma'}),
\end{equation}
where $n(E_{\veck,\sigma})$ is the Fermi-Dirac distribution at temperature $T$ and $n'(E_{\veck,\sigma})=-[4k_B T\cosh^2(E_{\veck,\sigma}/2k_BT)]^{-1}$.

The next step is to average out the $\gamma$--fermions.
The average $\langle \gamma_{\veck, \sigma}\gamma_{\veck',\sigma'}\rangle_{\gamma}=\langle \gamma^{\dagger}_{\veck, \sigma}\gamma^{\dagger}_{\veck',\sigma'}\rangle_{\gamma}=0$, and
\begin{equation}
    \langle \gamma^{\dagger}_{\veck, \sigma}\gamma_{\veck',\sigma'}\rangle_{\gamma}=\frac{\delta_{\veck,\veck'}\delta_{\sigma, \sigma'}}{1+\text{e}^{E_{\veck,\sigma}/k_BT}} = \delta_{\veck,\veck'}\delta_{\sigma, \sigma'}n(E_{\veck,\sigma}).
\end{equation} 
 We find the effective contribution to the interaction between the spins,
\begin{equation}\label{eq:effective spin-spin model p-wave}
    \begin{split}
        \mathcal{H}_\text{eff} =\sum_{i,j} \bigg(   J_{i,j} \vecS_i\cdot\vecS_j +   K_{i,j} S_i^z S_j^z +  D_{i,j}[\vecS_i\times \vecS_j]_z  +   L_{i,j}(S_i^x S_j^x - S_i^y S_j^y)+  M_{i,j}(S_i^x S_j^y + S_i^yS_j^x) \bigg).
    \end{split}
\end{equation}
The coupling constants are defined in Table \ref{tab:p-wave coupling constants}. \\

\begingroup
\setlength{\tabcolsep}{6pt} 
\renewcommand{\arraystretch}{2.5} 
\begin{table}[]
    \centering
    \begin{tabular}{|c|c|c|}
    \hline
   \textbf{Quantity} & \textbf{Definition} & \textbf{Properties} \\ \hline
      $ J_{i,j}$ & $\frac{\Lambda^2}{4N^2}\sum_{\vecq} A^{(1)}_{\vecq+\vecdQ,\downarrow,\uparrow}\cos( \vecq\cdot\vectau)$ & $J_{i,j}=J_{|\tau|}$ \\ \hline
      $   K_{i,j}$&$ \frac{\Lambda^2}{8N^2}\sum_{\vecq,\sigma} \bigg( A^{(1)}_{\vecq,\sigma,\sigma} + A^{(2)}_{\vecq,\sigma,\sigma} - A^{(1)}_{\vecq+\vecdQ,\downarrow,\uparrow}\bigg)\cos{(\vecq\cdot\vectau)}$ & $K_{i,j}=K_{|\tau|}$ \\ \hline
      $   L_{i,j} $&$ \frac{\Lambda^2}{4N^2}\cos\left(\vecdQ\cdot({r}_i+{r}_j)\right)\sum_{\vecq} A^{(2)}_{\vecq,\downarrow, \uparrow}\cos(\vecq\cdot\vectau)$ &- \\ \hline
      $   M_{i,j} $&$ -\frac{\Lambda^2}{4N^2}\sin\left(\vecdQ\cdot({r}_i+{r}_j)\right)\sum_{\vecq} A^{(2)}_{\vecq,\downarrow, \uparrow}\cos(\vecq\cdot\vectau)$ &  $M_{i,j}=0$ when $\Delta Q ^{\uparrow, \downarrow}=0$\\ \hline
         $     D_{i,j} $&$ \frac{\Lambda^2}{4N^2}\sum_{\vecq}A^{(1)}_{\vecq+\vecdQ,\downarrow,\uparrow}\sin(\vecq\cdot\vectau) $  & $D_{i,j}=\text{sgn}(\tau) D_{|\tau|}$, and $D_{i,j}=0$ when $\Delta Q ^{\uparrow, \downarrow}=0$ \\ \hline
     ${A}_{\vecq,\sigma, \sigma'}^{(1)} $&$ \sum_{\veck}\left( {F}_{\veck, \vecq, \sigma, \sigma'}^{(1)}+{F}_{\veck, \vecq, \sigma, \sigma'}^{(2)}+{F}_{\veck, \vecq, \sigma, \sigma'}^{(3)}+{F}_{\veck, \vecq, \sigma, \sigma'}^{(4)} \right) $ & ${A}_{\vecq,\sigma, \sigma'}^{(1)}={A}_{-\vecq,\sigma', \sigma}^{(1)}$  \\ \hline 
     ${A}_{\vecq,\sigma, \sigma'}^{(2)} $&$ \sum_{\veck}\left( {F}_{\veck, \vecq, \sigma, \sigma'}^{(5)}+{F}_{\veck, \vecq, \sigma, \sigma'}^{(6)}+{F}_{\veck, \vecq, \sigma, \sigma'}^{(7)}+{F}_{\veck, \vecq, \sigma, \sigma'}^{(8)} \right) $ & ${A}_{\vecq,\sigma, \sigma'}^{(2)}=({A}_{-\vecq,\sigma', \sigma}^{(2)})^*$ \\ \hline
     ${F}_{\veck, \vecq, \sigma, \sigma'}^{(1)} $&$ |u_{\veck+\vecq,\sigma}|^2|u_{\veck,\sigma'}|^2 \frac{n(\Tilde{E}_{\veck+\vecq,\sigma})-n(\Tilde{E}_{\veck,\sigma'})}{\Tilde{E}_{\veck+\vecq,\sigma}-\Tilde{E}_{\veck,\sigma'}}$ & ${F}_{\veck, \vecq, \sigma, \sigma'}^{(1)}={F}_{\veck+q, -\vecq, \sigma', \sigma}^{(1)}$ \\\hline
     ${F}_{\veck, \vecq, \sigma, \sigma'}^{(2)} $&$ |u_{\veck+\vecq,\sigma}|^2|v_{\veck,\sigma'}|^2 \frac{n(\Tilde{E}_{\veck+\vecq,\sigma})+n(\Tilde{E}_{-\veck,\sigma'})-1}{\Tilde{E}_{\veck+\vecq,\sigma}+\Tilde{E}_{-\veck,\sigma'}}$ & ${F}_{\veck, \vecq, \sigma, \sigma'}^{(2)}={F}_{\veck+q, -\vecq, \sigma', \sigma}^{(3)}$  \\ \hline
     ${F}_{\veck, \vecq, \sigma, \sigma'}^{(3)} $&$ |v_{\veck+\vecq,\sigma}|^2|u_{\veck,\sigma'}|^2 \frac{n(\Tilde{E}_{-\veck-\vecq,\sigma})+n(\Tilde{E}_{\veck,\sigma'})-1}{\Tilde{E}_{-\veck-\vecq,\sigma}+\Tilde{E}_{\veck,\sigma'}}$ & ${F}_{\veck, \vecq, \sigma, \sigma'}^{(3)}={F}_{\veck+q, -\vecq, \sigma', \sigma}^{(2)}$  \\ \hline
     ${F}_{\veck, \vecq, \sigma, \sigma'}^{(4)} $&$ |v_{\veck+\vecq,\sigma}|^2|v_{\veck,\sigma'}|^2 \frac{n(\Tilde{E}_{-\veck-\vecq,\sigma})-n(\Tilde{E}_{-\veck,\sigma'})}{\Tilde{E}_{-\veck-\vecq,\sigma}-\Tilde{E}_{-\veck,\sigma'}}$ & ${F}_{\veck, \vecq, \sigma, \sigma'}^{(4)}={F}_{\veck+q, -\vecq, \sigma', \sigma}^{(4)}$  \\ \hline 
      ${F}_{\veck, \vecq, \sigma, \sigma'}^{(5)} $&$ -u_{\veck+\vecq,\sigma}u^*_{\veck,\sigma'}v^*_{-\veck-\vecq,\sigma}v_{-\veck,\sigma'} \frac{n(\Tilde{E}_{\veck+\vecq,\sigma})-n(\Tilde{E}_{\veck,\sigma'})}{\Tilde{E}_{\veck+\vecq,\sigma}-\Tilde{E}_{\veck,\sigma'}}$ & ${F}_{\veck, \vecq, \sigma, \sigma'}^{(5)}=({F}_{\veck+q, -\vecq, \sigma', \sigma}^{(5)})^*$  \\ \hline
      ${F}_{\veck, \vecq, \sigma, \sigma'}^{(6)} $&$ -u_{\veck+\vecq,\sigma}v_{\veck,\sigma'}v^*_{-\veck-\vecq,\sigma}u^*_{-\veck,\sigma'} \frac{n(\Tilde{E}_{\veck+\vecq,\sigma})+n(\Tilde{E}_{-\veck,
      \sigma'})-1}{\Tilde{E}_{\veck+\vecq,\sigma}+\Tilde{E}_{-\veck,\sigma'}}$ & ${F}_{\veck, \vecq, \sigma, \sigma'}^{(6)}=({F}_{\veck+q, -\vecq, \sigma', \sigma}^{(7)})^*$ \\ \hline
      ${F}_{\veck, \vecq, \sigma, \sigma'}^{(7)} $&$ -v^*_{\veck+\vecq,\sigma}u^*_{\veck,\sigma'}u_{-\veck-\vecq,\sigma}v_{-\veck,\sigma'}  \frac{n(\Tilde{E}_{-\veck-\vecq,\sigma})+n(\Tilde{E}_{\veck,\sigma'})-1}{\Tilde{E}_{-\veck-\vecq,\sigma}+\Tilde{E}_{\veck,\sigma'}}$ & ${F}_{\veck, \vecq, \sigma, \sigma'}^{(7)}=({F}_{\veck+q, -\vecq, \sigma', \sigma}^{(6)})^*$ \\ \hline
      ${F}_{\veck, \vecq, \sigma, \sigma'}^{(8)} $&$ -v^*_{\veck+\vecq,\sigma}v_{\veck,\sigma'}u_{-\veck-\vecq,\sigma}u^*_{-\veck,\sigma'}   \frac{n(\Tilde{E}_{-\veck-\vecq,\sigma})-n(\Tilde{E}_{-\veck,\sigma'})}{\Tilde{E}_{-\veck-\vecq,\sigma}-\Tilde{E}_{-\veck,\sigma'}}$ & ${F}_{\veck, \vecq, \sigma, \sigma'}^{(8)}=({F}_{\veck+q, -\vecq, \sigma', \sigma}^{(8)})^*$ \\ \hline   
    \end{tabular}
    \caption{Coupling constants for the spin-spin interaction mediated by a superconductor with equal-spin Cooper pairs. In the cases where the denominator and nominator in $F^{(1)}, F^{(4)}, F^{(5)} \text{ and } F^{(8)}$ become zero, the limit is taken: $\text{lim}_{E_1\rightarrow E_2}\frac{n(E_1)-n(E_2)}{E_1-E_2} = -[4k_B T\cosh^2(E_2/2k_BT)]^{-1}$. }
    \label{tab:p-wave coupling constants}
\end{table}
\endgroup

\subsection{Numerical calculation of the coupling constants}
We calculate the coupling constants numerically at zero temperature. The coupling constants can also be calculated at higher temperatures, but the spin ground state of the adatoms is only defined at $T=0$. However, in the numerical procedure used to calculate the coupling constants in $\mathcal{H}_\text{eff}$, it is convenient to include a small temperature $T=0.01T_c$ well below the superconducting critical temperature $T_c$ to effectively turn the Fermi-Dirac distribution functions into slightly smeared step functions. This facilitates numerical convergence and avoids the requirement of computationally infeasible system sizes. 
The coupling constants are calculated on a lattice with $N=10\;000$ lattice points. We have confirmed that this size is sufficient for the coefficients to converge as a function of $N$.  
The exception is when the magnitude of the superconducting OP is small, because then the derivative of the Fermi-Dirac distribution is spiky and a larger number of lattice points is required. 

The coupling constants belonging to Fig. 1(c) and 2(c) in the main text are shown in Fig. \ref{fig:coupling constants spin current}. We observe that the coupling strength generally decreases when the distance $\tau=r_j-r_i$ increases. The coupling constants $L_{i,j}$ and $M_{i,j}$ depend on the absolute position $r_i$ of the spins. The numerically calculated coupling constants that were used to determine the ground state configuration for the other parameter variations (no supercurrent, charge supercurrent, different supercurrent strengths) are not shown here.

\begin{figure*}
    \centering
    \includegraphics{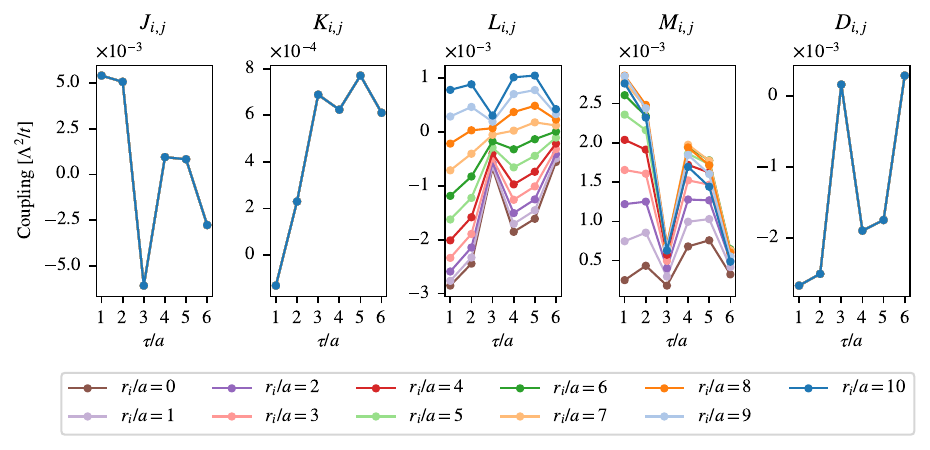}
    \caption{The coupling constants in units of $\Lambda^2/t$ for a superconductor with equal-spin Cooper pairs. The chemical potential is $\mu=-t$ and the magnitude of the superconducting OP is $\Delta^\sigma=0.1t$. A spin current $Q^\sigma/Q_c \approx 0.87\sigma$, or equivalently $aQ^{\sigma}=2\pi\sigma\cdot 7/1000$, flows through the superconductor. The coupling constants shown here were used to calculate the ground state configuration in Fig. 2(c) in the main text. 
    } 
    \label{fig:coupling constants spin current}
\end{figure*}

\subsection{Higher dimensions}
Our model is straightforwardly generalized to 2D or 3D by summing over all the components of the vectors $\boldsymbol{k}$ and $\boldsymbol{q}$ in the momentum space sums in Table \ref{tab:p-wave coupling constants}. Additionally, the expressions for $\epsilon_{\boldsymbol{k}}$ and $\Delta_{\boldsymbol{k}}^\sigma$ in any dimension are $\epsilon_{\boldsymbol{k}}=-\mu-t\sum_{\boldsymbol{\delta}} e^{i\boldsymbol{k}\cdot \boldsymbol{\delta}}$ and $\Delta_{\boldsymbol{k}}^\sigma=\sum_\delta \Delta_\delta^\sigma e^{i\boldsymbol{k}\cdot\boldsymbol{\delta}}$, where $\boldsymbol{\delta}$ refers to the vectors to all nearest neighbors. The main obstacle is that numerical calculations in 2D and 3D become extremely time-consuming since all the coupling constants contain terms of the kind $(n(E_1)-n(E_2))/(E_1-E_2)$. At low temperatures, this resembles a Dirac delta function for $\boldsymbol{k}$-values where $E_1\rightarrow E_2$, and a large system size is therefore needed to sample this part. For this reason, we have focused on the 1D case, but do not expect any qualitative changes to the effective spin-spin Hamiltonian in higher dimensions. The fact that our main results hold both in 2D and in 3D can be proven analytically as follows. Consider for instance the Dzyaloshinskii-Moriya interaction. We have
\begin{equation}
     D_{i,j}=\frac{\Lambda^2}{4N^2}\sum_{\boldsymbol{q}}A^{(1)}_{\boldsymbol{q}+\Delta\boldsymbol{Q}^{\uparrow\downarrow},\downarrow,\uparrow}\sin(\boldsymbol{q}\cdot\boldsymbol{\tau}), 
\end{equation}
where $\sum_{\boldsymbol{q}}$ now runs over a 2D or 3D Brillouin zone, $\Delta \boldsymbol{Q}^{\uparrow\downarrow}=\boldsymbol{Q}^\uparrow-\boldsymbol{Q}^\downarrow$, and in general $A^{(1)}_{\boldsymbol{q}, \downarrow, \uparrow} = A^{(1)}_{-\boldsymbol{q}, \uparrow, \downarrow}$. When the magnitude of the superconducting order parameter is spin independent,  $A^{(1)}_{\boldsymbol{q}, \downarrow, \uparrow} = A^{(1)}_{\boldsymbol{q}, \uparrow, \downarrow}$, so $A_{\boldsymbol{q}}^{(1)}$ is symmetric in $\boldsymbol{q}$. When $\Delta\boldsymbol{Q}^{\uparrow\downarrow}=0$, the DM coefficient is a sum over an even function times an odd function, which gives $D_{i,j}=0$.  When $\Delta\boldsymbol{Q}^{\uparrow\downarrow}\neq 0$, we have $A^{(1)}_{\boldsymbol{q}+\Delta\boldsymbol{Q^{\uparrow\downarrow}},\downarrow,\uparrow}\neq A^{(1)}_{-\boldsymbol{q}+\Delta\boldsymbol{Q}^{\uparrow\downarrow},\downarrow,\uparrow}$, so the DM term is nonzero. The larger the spin current is, the further $A^{(1)}_{\boldsymbol{q}+\Delta\boldsymbol{Q}^{\uparrow\downarrow},\downarrow,\uparrow}$ is shifted from being even and the larger $D_{i,j}$ becomes. The center-of-mass dependent coefficients $L_{i,j}$ and $M_{i,j}$ are given by
\begin{equation}
     L_{i,j} = \frac{\Lambda^2}{4N^2}\cos\left(\Delta\boldsymbol{Q}^{\uparrow\downarrow}\cdot(\boldsymbol{r}_i+\boldsymbol{r}_j)\right)\sum_{\boldsymbol{q}} A^{(2)}_{\boldsymbol{q},\downarrow, \uparrow}\cos(\boldsymbol{q}\cdot\boldsymbol{\tau})
\end{equation} 
\begin{equation}
     M_{i,j} = -\frac{\Lambda^2}{4N^2}\sin\left(\Delta\boldsymbol{Q}^{\uparrow\downarrow}\cdot(\boldsymbol{r}_i+\boldsymbol{r}_j)\right)\sum_{\boldsymbol{q}} A^{(2)}_{\boldsymbol{q},\downarrow, \uparrow}\cos(\boldsymbol{q}\cdot\boldsymbol{\tau}).
\end{equation}
Since the magnitude of the superconducting order parameter is taken to be spin independent and real in $\boldsymbol{k}$-space, we have $A^{(2)}_{\boldsymbol{q}, \downarrow, \uparrow} = A^{(2)}_{-\boldsymbol{q}, \downarrow, \uparrow}$. The sum is therefore a sum over two even functions in $\boldsymbol{q}$, which is generally nonzero. In fact, the $\boldsymbol{q}$-sum is identical to the $\boldsymbol{q}$-sum in the anisotropic exchange coupling $K_{i,j}$ when the magnitude of the superconducting order parameter is spin independent and real in $\boldsymbol{k}$-space. In the absence of a spin supercurrent, $M_{i,j}=0$ while $L_{i,j}\neq 0$, although $L_{i,j}$ loses its center-of-mass dependence. In the presence of a spin supercurrent, both coupling constants are generally nonzero and dependent on the center-of-mass coordinate regardless of the dimensionality of the system.

\section{Calculating the ground state configuration}

The ground state configuration of the spins $\vecS_i$ is found by minimizing $\mathcal{H}_\text{eff}$. Upon increasing the temperature, the probability for the system to be in any spin configuration $\{\vecS_i \}$ is given by the Boltzmann factor $\text{e}^{-\beta\mathcal{H}_\text{eff}(\{\vecS_i \})}/Z$, where $Z$ is the partition function. Since our effective Hamiltonian is derived using perturbation theory, the coupling constant $\Lambda$ and thus $\mathcal{H}_\text{eff}$ are small. Therefore, at a finite temperature that exceeds the magnitude of the highest excited state of the Hamiltonian, all spin configurations are sampled in the partition function. However, a non-perturbative approach would in all likelihood yield larger coefficients in $\mathcal{H}_\text{eff}$, indicating that the spins would stay close to their ground state configuration upon increasing the temperature, thus allowing for the possibility to design a spin lattice at finite temperature. In other words, the stronger the coupling, the more robust the ground state spin configuration against thermal fluctuations.

The first step in calculating the ground state configuration is to calculate the coefficients in $\mathcal{H}_\text{eff}$ at zero temperature. It is convenient to include a small temperature $T=0.01T_c$ in the numerical procedure used to calculate the coupling constants in Eq. \eqref{eq:effective spin-spin model p-wave}, because this effectively turns the Fermi-Dirac distribution functions into slightly smeared step functions.
When the coupling constants are calculated, it is possible to calculate the ground state of the spins. To demonstrate how the ground state is altered by the supercurrent, we consider two spins $\vecS_1$ and $\vecS_2$ placed on the superconductor. 
The spins can take the values
\begin{equation}
    \vecS_1,\vecS_2 = \cos \vartheta \sin \varphi \boldsymbol{e_x} + \sin \vartheta \sin \varphi \boldsymbol{e_y}+\cos \varphi \boldsymbol{e_z}. 
\end{equation}
The ground state configuration is found "brute force" by calculating $\mathcal{H}_\text{eff}$ for $M$ equally spaced values for $\varphi\in[0, 2\pi)$ and $M/2$ equally spaced values for $\theta\in [0,\pi)$, and finding the configuration that gives the lowest value for $\mathcal{H}_\text{eff}$. This gives in total $M^4/4$ combinations for the spins $\vecS_1$ and $\vecS_2$. $M=360$ gives a 1 degree uncertainty, and this is the value we have used when determining the ground state of the dimers (two spins). When we calculate the ground state of trimers (three spins), we use $M=60$ which gives an uncertainty of 6 degrees. The reduced accuracy for the trimers is due to the increased computational cost of calculating the ground state for three spins.

The ground state configuration remains non-collinear and center-of-mass dependent, as in the main manuscript, when the chemical potential is increased or when the spin current magnitude is decreased. This is shown in Fig. \ref{fig:parameter-variations-of-gs}. In Fig. \ref{fig:gs currentdependence}, the ground state dependence on the spin and charge supercurrent magnitude is shown. We see that the ground state configuration can be tuned by a supercurrent.
Upon decreasing the magnitude of the superconducting OP, the coupling constant $J_{i,j}$ becomes dominant and the ground state configuration becomes approximately collinear without specific preferences for the axis alignment. Since the effective Hamiltonian contains five different types of interactions which are all generally non-zero in the presence of a spin supercurrent, there are competing interactions in the Hamiltonian which cannot be minimized simultaneously. This is exactly the definition of magnetic frustration. Therefore, we expect that the magnetic ground state is in general frustrated and that the spin supercurrent does tune the frustration by altering the relative magnitude of the coefficients.
\begin{figure}
    \centering
    \includegraphics{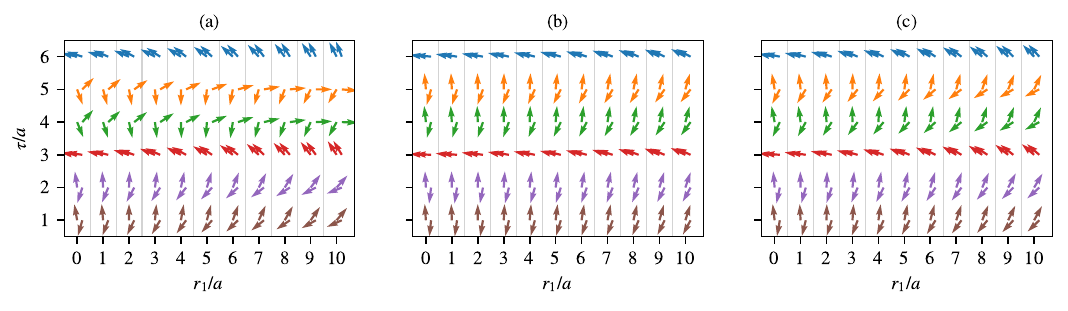}
    \caption{The ground state configuration in the $xy$-plane of two spins $\vecS_1$ and $\vecS_2$ whose interaction is mediated by a superconductor with equal-spin Cooper pairs. The parameters used are the same as in Fig. 2(c) in the main text, except (a) $\mu/t=+1$, (b) $Q^\sigma/Q_c\approx 0.37 \sigma$ 
    and (c) $Q^\sigma/Q_c\approx 0.62 \sigma$. 
    }
    \label{fig:parameter-variations-of-gs}
\end{figure}

\begin{figure}
    \centering
    \includegraphics{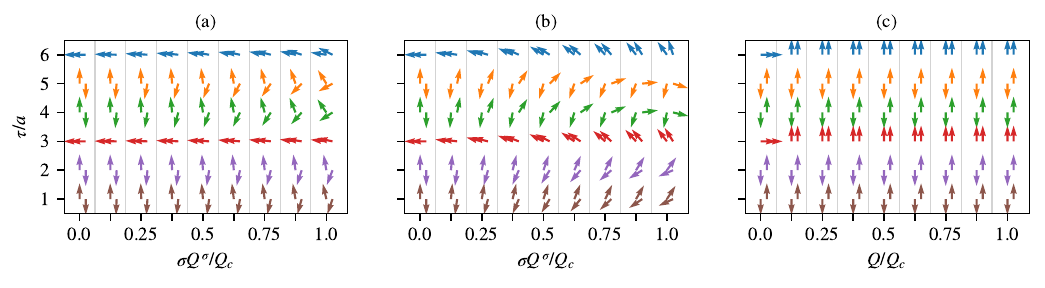}
    \caption{(a-b): Spin supercurrent dependence on the magnetic ground state.  The position of $\boldsymbol{S}_1$ is $r_1/a=0$ in (a) and $r_1/a=8$ in (b). (c) Charge supercurrent dependence on the magnetic ground state. The ground state at $Q=0$ is degenerate at $\tau/a=3$ and 6, but the energy is not minimized for the ferromagnetic alignment along $y$.}
    \label{fig:gs currentdependence}
\end{figure}

Figure \ref{fig:trimer} shows the magnetic ground state for spin trimers
with different spin positions and supercurrents in the superconductor. Importantly, the results from the dimer calculation hold even for the trimers. First, we see that we get non-collinear ground states. In the absence of a spin supercurrent, this non-collinearity arises due to antiferromagnetic next-nearest neighbor coupling with almost the same strength as the nearest neighbor antiferromagnetic coupling. When the spin supercurrent is turned on, we get DM interaction between the spins as an additional source of spin non-collinearity. Second, the ground state of the trimer is center-of-mass dependent in the presence of a spin supercurrent. Therefore, the magnetic ground state of the impurity spins can be tuned by placing the spins appropriately on the surface. Moreover, the ground state of the trimer can be tuned with the magnitude of the spin supercurrent. This is shown in Fig. 2(e) in the main text.
\begin{figure}
    \centering
    \includegraphics{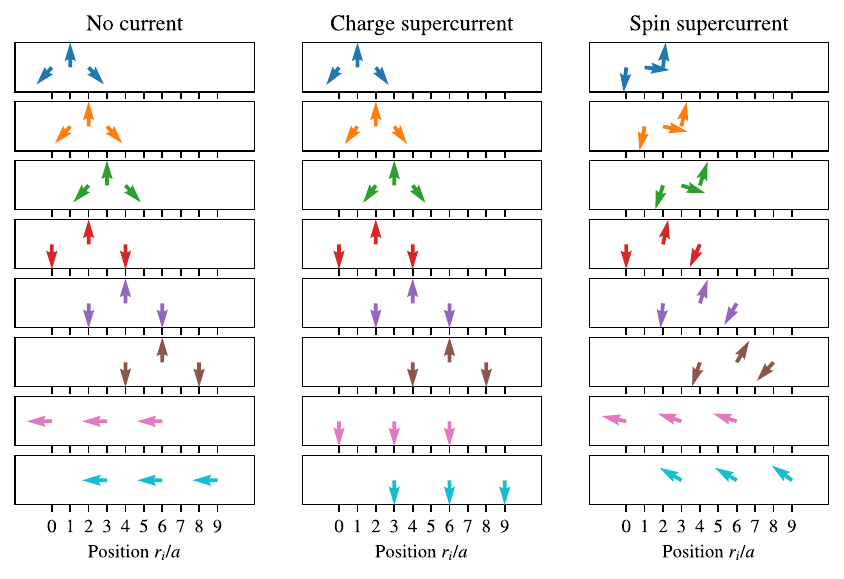}
    \caption{Ground states for spin trimers with the three spins placed at various positions on top of the superconductor. The spins are in the $xy$-plane, and the position of each of the three spins is shown along the horizontal axis. The magnitude of the charge supercurrent is given by $Q^{\sigma}/Q_c\approx 0.87$, and the magnitude of the spin supercurrent is given by $Q^{\sigma}/Q_c\approx 0.87 \sigma$ as in the main paper. The accuracy of the spin orientations is 6 degrees (limited by a high numerical computation cost).}
    \label{fig:trimer}
\end{figure}

\section{The origin for the center-of-mass coordinate of the magnetic adatoms}
The physical quantity that defines the origin of the axis from which the position $r_i$ is measured is the phase difference between the superconducting order parameters describing Cooper pairs with opposite spin.  The superconducting order parameter for spin-$\sigma$ Cooper pairs is
\begin{equation}\label{eq:espdelta}
    \Delta_{i,j}^{\sigma}= \Delta_{\delta}^{\sigma} \text{e}^{\text{i} Q^{\sigma}\cdot({r}_i+{r}_j)},
\end{equation}
where $i,j$ are nearest neighbors. Now, consider $r_i<r_j$ without loss of generality. The phase difference between spin up and spin down Cooper pairs is then $\Delta Q(2r_i + a)$. The origin $r=0$ is the point where the phase difference between the Cooper pairs is $\Delta Q a$ when the electrons in the Cooper pairs are located at this lattice site and the one to the right of it. For $r_j<r_i$, the order parameter picks up a phase $\pi$ due to $\Delta_\delta^\sigma$ switching sign, but this does not affect the phase difference between the order parameters describing Cooper pairs with opposite spin. 
Since the phase-winding causes the supercurrent-carrying superconducting order parameters $\Delta^\sigma_{i,j}$ to be periodic in space, there are several equivalent positions that may be used to define the point $r_i=0$.
Measuring this phase difference directly would be difficult, but it nevertheless has an experimentally measurable consequence: the center-of-mass dependence of the spin interactions predicted in our work as seen by experimentally by placing several pairs of spins at an arbitrary position on the surface of a superconductor. The magnetic ground-state of the spin pairs will then in general be different precisely due to the center-of-mass dependence induced by the spin supercurrent in our system. 

\section{Magnon gap}
\subsection{Equal spin pairing}
The first-order contribution to the magnon spectrum is found by calculating 
\begin{equation}
\begin{split}
   \langle H_c\rangle_{\gamma}=\sum_{\sigma, \sigma'}\sum_{\veck, \vecq} T_{q,\sigma,\sigma'}\langle c^\dagger_{\veck+\vecq,\sigma}c_{\veck, \sigma'}\rangle_{\gamma}, \quad T_{q,\sigma\sigma'}=\sum_{l\in\{A,B\}}\frac{\Lambda_{{l}}}{{2}N}\sum_{i{\in l}}\vecS_i\cdot\boldsymbol{\sigma}_{\sigma, \sigma'}\text{e}^{-\text{i}\vecq\cdot{r}_i},
\end{split}
\end{equation}
where $A,B$ denote the sublattices of the antiferromagnet. We assume for simplicity that the $s-d$-coupling strength $\Lambda_l=\Lambda$ is the same for both sublattices.
For a superconductor with equal-spin Cooper pairs, the coherence factors $u_{k,\sigma}$ and $v_{k,\sigma}$ that diagonalize the Hamiltonian are given in Eq. \eqref{eq:coherence factors equal spin pairing}. The corresponding eigenvalues $E_{k,\sigma}$ are given in Eq. \eqref{eq:eigenvalues equal spin pairing}. Rewriting the $c$--operators in terms of the $\gamma$--operators according to Eq. \eqref{eq:bog transf equal spin pairing} and evaluating the ground-state quantum expectation value for the $\gamma$--operators gives
\begin{equation}
    \langle c^\dagger_{\veck+\vecq,\sigma}c_{\veck, \sigma'}\rangle_{\gamma} = \delta_{q,0}\delta_{\sigma,\sigma'}\left( |u_{k-Q^{\sigma},\sigma}|^2 n(E_{k,\sigma}) + |v_{k-Q^{\sigma},\sigma}|^2 [1- n(E_{-k+2Q^{\sigma},\sigma}) ] \right).
\end{equation}
The effective contribution to the spin Hamiltonian is
\begin{equation}
\label{eq:expectation value of Hc}
    \langle H_c \rangle_{\gamma} = \sum_{l\in \{A,B\}}\sum_{i\in l} S_i^z\cdot\left( \frac{\Lambda}{2N}\sum_{k,\sigma}\sigma \left[ |u_{k-Q^{\sigma},\sigma}|^2 n(E_{k,\sigma}) + |v_{k-Q^{\sigma},\sigma}|^2 (1- n(E_{-k+2Q^{\sigma},\sigma})) \right] \right) =  -\sum_{l\in \{A,B\}}\sum_{i\in l} S_i^z\cdot \mathcal{D}
\end{equation}
with
\begin{equation}\label{eq:D for equal spin pairing}
    \mathcal{D}= -\frac{\Lambda}{2N} \sum_{k,\sigma} \sigma  \big( |u_{k,\sigma}|^2 n(E_{k+Q^{\sigma},\sigma}) + |v_{k,\sigma}|^2 [1- n(E_{-k+Q^{\sigma},\sigma}) ] \big).
\end{equation}
This is effectively an applied magnetic field $\mathcal{D}$ in the $z$-direction.
This connection can also be seen directly by calculating the magnetization $M^z$ in the superconductor,
\begin{equation}
    \langle M^z \rangle_\gamma = \bigg\langle \frac{1}{N} \sum_{k,\sigma}\sigma c_{k,\sigma}^\dagger c_{k,\sigma}\bigg\rangle_\gamma = -\frac{2}{\Lambda} \mathcal{D}.
\end{equation}
The physical interpretation is that when the spin-$\sigma$ Cooper pairs gain a momentum $Q^\sigma$, the occupation of the spin-$\sigma$ quasiparticles changes and the superconductor gains a net magnetization. The magnetization is independent of the sign of the Cooper pair momentum $Q^\sigma$.

The magnon gap is found by diagonalizing the following Hamiltonian, which, depending on the interaction parameter choices, can model either a conventional antiferromagnet or altermagnet \cite{hodt2024spin}:
\begin{equation}
    H_\text{AM} = \sum_{\langle i,j \rangle} J_1 \vecS_i\cdot \vecS_j - \sum_{i} [K_1(S_i^z)^2 + \mathcal{D}S_i^z] + \bigg( J_2 \bigg[ \sum_{\langle\langle i_x,j_x\rangle \rangle\in A} + \sum_{\langle\langle i_y,j_y\rangle \rangle\in B} \bigg] + J_2' \bigg[ \sum_{\langle\langle i_y,j_y\rangle \rangle\in A} +\sum_{\langle\langle i_x,j_x\rangle \rangle\in B} \bigg] \bigg) \vecS_i\cdot \vecS_j.
\end{equation}
The model is illustrated in Fig. \ref{fig:altermagnet}. $\langle \langle i_d,j_d \rangle \rangle \in l$ denotes the pairs of next-nearest neighbors connected along the $d$-direction within the $l$ sublattice. This altermagnetic model consists of two superimposed square lattices, where the interactions are a nearest-neighbor interaction $J_1>0$, an anisotropic next-nearest-neighbor interaction $J_2,J_2'$ which is rotated for one sublattice compared to the other, and an anisotropy term $K_1<0$. A pure antiferromagnetic model is achieved by setting $J_2=J_2'=0$.

\def\ptsSublatticeA{(0,0), (2,0), (4,0), (2,2), (4,2),     (6, 0), (6, 2)}
\def\ptsSublatticeB{(0+1,0+1), (2+1,0+1), (4+1,0+1), (2+1,2+1), (4+1,2+1),
(6+1, 0+1), (6+1, 2+1)}

\begin{figure}
\begin{tikzpicture}[
dotA/.style = {circle, fill=blue!80, minimum size=6pt, inner sep=0pt},
dotB/.style = {circle, fill=red!80, minimum size=6pt, inner sep=0pt},
dotMBZ/.style = {circle, fill=black!100, minimum size=6pt, inner sep=0pt}
                    ]

\foreach \q [count=\l] in \ptsSublatticeA 
    \node (n\l) [dotA] at \q {};
\foreach \q [count=\l] in \ptsSublatticeB
    \node (n\l) [dotB] at \q {};
\node (n) [dotA, label={[above right, color=blue]:$B$}] at (0,2) {};
\node (n) [dotB, label={[above right, color=red]:$A$}] at (1,3) {};

\draw[thick, dotted] (0,0) --  node[above] {$J_1$} ++(1,1);

\draw[blue, thick, dotted] (2,0) --  node[left] {$J_2$} ++(0,2);
\draw[blue, thick, dotted] (2,0) --  node[below] {$J_2'$} ++(2,0);

\draw[red, thick, dotted] (3,1) --  node[left] {$J_2'$} ++(0,2);
\draw[red, thick, dotted] (3,1) --  node[below] {$J_2$} ++(2,0);

\draw[->, thick] (-0.5,-0.5) -- (0.5,-0.5) node[right] {$x$};
\draw[->, thick] (-0.5,-0.5) -- (-0.5,0.5) node[above] {$y$};

\draw[decorate,decoration={brace,amplitude=5pt,raise=2.5pt},yshift=0pt] (5,3) -- (7,3) node [midway,yshift=13pt, xshift =0pt]{$a$};    
\end{tikzpicture}
\caption{Altermagnetic model.}
\label{fig:altermagnet}
\end{figure}

We apply the Holstein-Primakoff transformation to the spin operators on each sublattice:
\begin{equation}
\begin{split}
    S_{i\in A}^z = S-  a_i^\dagger a_i, \qquad S_{i\in A}^+ \approx \sqrt{2S}a_i, \qquad S_{i\in A}^- \approx \sqrt{2S}a_i^\dagger \\  S_{i\in B}^z = -S+  b_i^\dagger b_i, \qquad S_{i\in B}^+ \approx \sqrt{2S}b_i^\dagger, \qquad S_{i\in B}^- \approx \sqrt{2S}b_i. 
\end{split}
\end{equation}
We then Fourier transform $a_k = \sqrt{2/N}\sum_{i\in A}a_i \text{e}^{\text{i}kr_i}$ and $b_k = \sqrt{2/N}\sum_{i\in B}b_i \text{e}^{\text{i}kr_i}$ and find
\begin{equation}
    H_\text{AM}=\sum_{k\in\text{MBZ}} \begin{pmatrix} a_k^\dagger & b_{-k}\end{pmatrix} \begin{pmatrix}
        J^* +S(J_2\gamma_k^x +J_2'\gamma_k^y)+\mathcal{D} & J_1S\gamma_k \\ J_1 S\gamma_k & J^* + S(J_2'\gamma_k^x + J_2 \gamma_k^y)-\mathcal{D}
    \end{pmatrix} \begin{pmatrix}a_k \\ b_{-k}^\dagger \end{pmatrix},
\end{equation}
where MBZ is the magnetic Brillouin zone, $\gamma_k = 4\cos(ak_x/2)\cos(ak_y/2)$, $\gamma_k^x=2\cos(ak_x)$, $\gamma_k^y=2\cos(ak_y)$ and $J^* = S(4J_1+2K_1-2(J_2+J_2'))$. The Hamiltonian is diagonalized through the Bogoliubov transformation 
\begin{equation}
    \begin{pmatrix}
        \alpha_k \\ \beta_{-k}^\dagger 
    \end{pmatrix} = \begin{pmatrix}
        \mu_k & \nu_k \\ \nu^*_k & \mu_k^*
    \end{pmatrix} \begin{pmatrix}
        a_k \\ b_{-k}^\dagger
    \end{pmatrix},
\end{equation}
and $|\mu_k|^2-|\nu_k|^2=1$ for the new operators to be bosonic. The diagonalized Hamiltonian becomes 
\begin{equation}
H_\text{AM} = \sum_{k\in \text{MBZ}} (\omega_{\alpha,k} \alpha_k^\dagger \alpha_k + \omega_{\beta,k}\beta_k^\dagger\beta_k), 
\end{equation}
with eigenvalues
\begin{equation}
    \omega_{\alpha,k} = \frac{H_k^{11}-H_k^{22}}{2} + \frac{1}{2}\sqrt{(H_k^{11}+H_k^{22})^2 - 4(SJ_1\gamma_k)^2}, \qquad   \omega_{\beta,k} = \frac{H_k^{22}-H_k^{11}}{2} + \frac{1}{2}\sqrt{(H_k^{11}+H_k^{22})^2 - 4(SJ_1\gamma_k)^2}
\end{equation}
and coherence factors
\begin{equation}
    \mu_k = \sqrt{\frac{1}{2} + \frac{1}{2}\frac{H_k^{11}+H_k^{22}}{\sqrt{(H_k^{11}+H_k^{22})^2 -4(J_1S\gamma_k)^2}}}, \qquad \nu_k = \text{sgn}(J_1S\gamma_k(H_k^{11}+H_k^{22}))\sqrt{-\frac{1}{2}+\frac{1}{2 }\frac{H_k^{11}+H_k^{22}}{\sqrt{(H_k^{11}+H_k^{22})^2 -4(J_1S\gamma_k)^2}}}.
\end{equation}
Here, $H_k^{11}= J^* +S(J_2\gamma_k^x +J_2'\gamma_k^y)+\mathcal{D} $ and $H_k^{22}=J^* + S(J_2'\gamma_k^x + J_2 \gamma_k^y)-\mathcal{D}$. We see that the eigenvalues are $\omega_{\alpha,k}=\omega'_{\alpha,k} +\mathcal{D}$ and $\omega_{\beta,k}=\omega'_{\beta,k} -\mathcal{D}$, where $\omega'_{\alpha,k}$ and $\omega'_{\beta,k}$ do not depend on the supercurrent. This shows that $\mathcal{D}$ is the contribution to the magnon gap from the supercurrent, and that the magnon gap is enhanced for magnon species $\alpha$ and suppressed for magnon species $\beta$ when $\mathcal{D}>0$.

If the triplet superconducting OP has equal magnitude for both spins, $|\Delta_k^{\sigma}|=|\Delta_k^{-\sigma}|$, the magnon gap correction $\mathcal{D}$ is zero in the presence of a pure charge supercurrent or a pure spin supercurrent. For the charge current $Q^{\sigma}=Q$, the absolute value of the coherence factors in Eq. \eqref{eq:coherence factors equal spin pairing} and the eigenenergies in Eq. \eqref{eq:eigenvalues equal spin pairing} are spin independent. Therefore, the spin sum in Eq. \eqref{eq:D for equal spin pairing} is zero. For a pure spin supercurrent $Q^{\sigma}=-Q^{-\sigma}$, we use that $\epsilon_k=\epsilon_{-k}$. Again, we find that the absolute value of the coherence factors in Eq. \eqref{eq:coherence factors equal spin pairing} are spin independent, and the eigenenergy $E_{k,\sigma}=E_{-k,-\sigma.}$. Since $|u_{k}|=|u_{-k}|$ and $|v_{k}|=|v_{-k}|$, we find again that the spin sum and thus $\mathcal{D}$ is zero. A non-zero $\mathcal{D}$ is thus obtained for a spin-polarized charge supercurrent ($Q^{\uparrow}\neq \pm Q^{\downarrow}$).

\subsection{Spin singlet $s$-wave Cooper pairs}

The attractive Hubbard Hamiltonian in real space for a spin-split spin singlet $s$-wave superconductor is
\begin{equation}
\begin{split}
    H_\text{SC}^s =\sum_{i,\sigma}(-\mu-\sigma h)c_{i,\sigma}^{\dagger} c_{i,\sigma} -2t \sum_{\langle i,j\rangle,\sigma} c_{i,\sigma}^{\dagger} c_{j,\sigma}-\frac{1}{8}U \sum_{ i}\sum_{\sigma} c_{i,\sigma}^{\dagger} c_{i,\sigma}c_{i,-\sigma}^{\dagger} c_{i,-\sigma}.
\end{split}
\end{equation}
Here, $h$ is the spin splitting. We apply the mean-field approximation to the last term in $H_\text{SC}^s$. The real-space order parameter is defined as $\Delta_{i}^{\sigma,-\sigma}=-U\langle c_{i,-\sigma}c_{i,\sigma}\rangle/4$, and we assume that it is an $s$-wave superconductor and the OP takes the form 
\begin{equation}
    \Delta_{i}^{\sigma,-\sigma}=  \Delta^{\sigma,-\sigma} \text{e}^{\text{i}2\vecQ\cdot{r}_i},
\end{equation}
The momentum $Q$ of the Cooper pairs is now spin-independent.
Next, we Fourier transform the electron operators. This gives the Hamiltonian
\begin{equation}
\begin{split}
    H_\text{SC}^s = \sum_{\veck, \sigma} (\epsilon_{\veck}-\sigma h) c_{\veck, \sigma}^{\dagger} c_{\veck, \sigma} + \frac{1}{2}\sum_{\veck, \sigma}(\Delta^{\sigma,-\sigma})^{\dagger}c_{-\veck+\vecQ,-\sigma}c_{\veck+\vecQ, \sigma} \text{+h.c.} 
\end{split}
\end{equation}
up to a constant. The superconducting OP satisfies $\Delta^{\uparrow,\downarrow}=-\Delta^{\downarrow,\uparrow}=\Delta$.
In the basis $\phi_{\veck,\sigma}=\begin{pmatrix}c_{\veck+\vecQ, \sigma} & c^{\dagger}_{-\veck+\vecQ,-\sigma}\end{pmatrix}^T$, the Hamiltonian is
\begin{equation}
    H_\text{SC}^s = \frac{1}{2}\sum_{\veck, \sigma}\phi^\dagger_{\veck, \sigma}\begin{pmatrix}
        \epsilon_{\veck+\vecQ}-\sigma h & \Delta^{\sigma,-\sigma} \\ (\Delta^{\sigma,-\sigma})^* & -\epsilon_{-\veck+\vecQ}-\sigma h
    \end{pmatrix}\phi_{\veck, \sigma}.
\end{equation}
Diagonalization is achieved through the Bogoliubov transformation 
\begin{equation}\label{eq:bog transf singlet}
    \begin{pmatrix}\gamma_{\veck+\vecQ, \sigma} \\ \gamma^{\dagger}_{-\veck+\vecQ,-\sigma}\end{pmatrix}=\begin{pmatrix}
        u^s_{\veck, \sigma} & v^s_{\veck, \sigma} \\ -v^{s*}_{\veck, \sigma} & u^{s*}_{\veck, \sigma}
    \end{pmatrix}\begin{pmatrix}c_{\veck+\vecQ, \sigma} \\ c^{\dagger}_{-\veck+\vecQ,-\sigma}\end{pmatrix},
\end{equation}
where $|u^s_{\veck,\sigma}|^2+|v^s_{\veck,\sigma}|^2=1$ for the $\gamma$-operators to be fermionic, and $u^s_{\veck,\sigma}=u^s_{-\veck,-\sigma}$ and $v^s_{\veck,\sigma}=-v^s_{-\veck,-\sigma}$ for the transformation to be consistent. The diagonalized Hamiltonian is $H_\text{SC}^s=\sum_{k,\sigma} E^s_{k,\sigma} \gamma_{k,\sigma}^{\dagger}\gamma_{k,\sigma}$. This gives the coherence factors
\begin{equation}
    u^s_{\veck,\sigma}=\sqrt{\frac{1}{2}+\frac{1}{2}\frac{\epsilon_{\veck+\vecQ}+\epsilon_{-\veck+\vecQ}}{\sqrt{(\epsilon_{\veck+\vecQ}+\epsilon_{-\veck+\vecQ})^2+4|\Delta|^2}}},
    \quad
    v^s_{\veck,\sigma}=\frac{\Delta^{\sigma,-\sigma}}{|\Delta^{\sigma,-\sigma}|} \sqrt{\frac{1}{2}-\frac{1}{2}\frac{\epsilon_{\veck+\vecQ}+\epsilon_{-\veck+\vecQ}}{\sqrt{(\epsilon_{\veck+\vecQ}+\epsilon_{-\veck+\vecQ})^2+4|\Delta|^2}}},
\end{equation}
and the eigenvalues
\begin{equation}
\begin{split}
    E^s_{\veck, \sigma}&=\frac{\epsilon_{\veck}-\epsilon_{-\veck+2\vecQ}}{2} + \sqrt{\bigg(\frac{\epsilon_{\veck}+\epsilon_{-\veck+2\vecQ}}{2}\bigg)^2+ |\Delta|^2} -\sigma h.
\end{split}
\end{equation}
This gives, as before, magnon energies $\omega_{\alpha,k}=\omega'_{\alpha,k} +\mathcal{D}^s$ and $\omega_{\beta,k}=\omega'_{\beta,k} -\mathcal{D}^s$, where $\omega'_{\alpha,k}$ and $\omega'_{\beta,k}$ do not depend on the supercurrent. The magnon gap correction is
\begin{equation}
    \mathcal{D}^s= -\frac{\Lambda}{2N} \sum_{k,\sigma} \sigma  \left( |u^s_{k-Q,\sigma}|^2 n(E^s_{k,\sigma}) + |v^s_{k-Q,\sigma}|^2 [1- n(E^s_{-k+2Q,-\sigma}) ] \right).
\end{equation}
The magnitudes of the coherence factors are spin-independent and the eigenenergy depends on spin only through the spin splitting field. Therefore, a non-zero spin splitting is required to achieve a nonzero magnon gap correction. The magnon gap correction can be simplified by shifting $\sum_k \rightarrow \sum_{k+Q}$ and relabeling $k,\sigma \rightarrow -k,-\sigma$ in the sum over the $v$--term:
\begin{equation}\label{eq:D for opposite spin pairing}
    \mathcal{D}^s= -\frac{\Lambda}{2N} \sum_{k,\sigma} \sigma n(E^s_{k,\sigma}).
\end{equation}

\section{Gap equation for the superconducting order parameter}

In this section, we identify the critical supercurrent magnitude $Q_c$ from solving the gap equation. This value is used to normalize the supercurrent strengths in the figures in the main text and in the Supplemental material. Moreover, we solve the gap equation to determine how the magnitude $\Delta^\sigma$ of the superconducting order parameter depends on the applied supercurrent, the spin splitting and the temperature. The resulting value is used in the calculations for the coupling constants in Eq. \eqref{eq:effective spin-spin model p-wave} and in the magnon gap corrections in Eqs. \eqref{eq:D for equal spin pairing} and \eqref{eq:D for opposite spin pairing}.\\

We underline that we consider an effective 1D model for simplicity since this allows us to numerically simulate larger systems, whereas the actual experimental realization of our setup would be in 3D (a thin-film geometry). It is well-known that effective 1D models can describe experimental systems in 3D well when physical quantities primarily vary along one coordinate axis \cite{lesueur_prl_08}.

\subsection{Equal spin pairing}
The superconducting order parameter is defined as $\Delta_{i,j}^\sigma=-U\langle c_{j,\sigma}c_{i,\sigma}\rangle/4 = \Delta_\delta^\sigma \text{e}^{\text{i}Q^\sigma(r_i+r_j)}$, where $i$ and $j$ are nearest neighbors. Moreover, the OP in $k$-space was defined as $\Delta_k^\sigma = 2i\Delta_a^\sigma \sin(ak)=\Delta^\sigma \sin(ak)$. Note that we keep the spin dependence of the OP here for generality. Rearranging the above definitions gives
\begin{equation}
    \Delta^\sigma = 2\text{i}\left(-\frac{U}{4} \langle c_{i+a,\sigma}c_{i,\sigma}\rangle \text{e}^{-\text{i}Q^\sigma(2r_i+a)} \right).
\end{equation}
We Fourier transform the electron operators and perform a sum over all lattice points to remove the $r_i$-dependence. This gives
\begin{equation}
    \Delta^\sigma = \frac{U}{2N}\sum_k \langle c_{k+Q^\sigma,\sigma}c_{- k+Q^\sigma,\sigma}\rangle\sin(ak). 
\end{equation}
The electron operators are replaced with $\gamma$-operators according to Eq. \eqref{eq:bog transf equal spin pairing}. 
The gap equation becomes
\begin{equation}
    \Delta^\sigma = -\frac{U}{4N} \sum_k \frac{\Delta^\sigma\sin^2(ak)}{\sqrt{(\epsilon_{k+Q^\sigma}+\epsilon_{-k+Q^\sigma})^2/4+|\Delta^\sigma\sin(ak)|^2}} \left[ n(E_{k+Q^\sigma,\sigma}) +n(E_{-k+Q^\sigma,\sigma})-1 \right].
\end{equation}
Figure \ref{fig:gapeq triplet}(a) shows the $Q^\sigma$-dependence of the OP magnitude. Below the critical current, it is reasonable to approximate the gap as $Q^\sigma$-independent. Therefore, it is reasonable to approximate the OP magnitude as spin independent since the OP magnitude depends on spin only through the Cooper pair momentum $Q^\sigma$.  The critical Cooper pair momentum is $aQ_c\approx 0.0505 \approx 2\pi\cdot 8/1000$ at $T=0$ and $aQ_c\approx 0.035 \approx 2\pi\cdot 6/1000$ at $T=0.2T_c$. Here, $T_c\approx 0.05t/k_B$ is the critical temperature when $Q^\sigma=0$, as shown in Fig. \ref{fig:gapeq triplet}(b). The supercurrent magnitude $aQ\approx 0.044$ used in Fig. 1 and 2 in the main text is below the critical supercurrent at $T=0$. 
\begin{figure}
    \centering
    \includegraphics[]{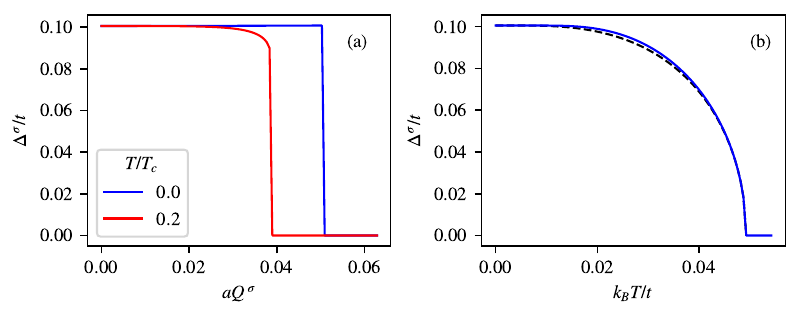}
    \caption{Numerical solution to the gap equation for the spin triplet superconductor. The chemical potential is $\mu/t=-1$ and $U/t=4.1$. In (b), $Q^\sigma=0$. The dashed black line is the approximate solution $\Delta^\sigma = \Delta_0 \tanh \left(1.74\sqrt{T_c/T-1  }\right)$ \cite{sigrist2005introduction}.
    }
    \label{fig:gapeq triplet}
\end{figure}

\subsection{Spin singlet $s$-wave Cooper pairs}%
The superconducting OP for a spin singlet $s$-wave superconductor is defined as 
\begin{equation}
    \Delta_i^{\uparrow,\downarrow} = -\frac{U}{4} \langle c_{i\downarrow}c_{i\uparrow}\rangle = \Delta \text{e}^{\text{i}2Qr_i}.
\end{equation}
We Fourier transform the electron operators,  and the gap equation takes the form
\begin{equation}
    \Delta = -\frac{U}{4}\sum_{k,k'}\frac{1}{N} \langle c_{k'\downarrow}c_{k\uparrow} \rangle\text{e}^{\text{i}(k+k'-2Q)r_i}.
\end{equation}
The $r_i$-dependence on the right-hand side is removed by performing the sum $\sum_i$ on both sides. This gives
\begin{equation}
    \Delta = -\frac{U}{4N}\sum_k \langle c_{-k+Q,\downarrow} c_{k+Q,\uparrow} \rangle.
\end{equation}
We use the transformation in Eq. \eqref{eq:bog transf singlet} and find
\begin{equation}
    \Delta = -\frac{U}{4N} \sum_k u^{s*}_{k\uparrow}v^s_{k\uparrow}\left[ n(E^s_{k+Q,\uparrow}) +n(E^s_{-k+Q,\downarrow})-1 \right].
\end{equation}
Inserting the coherence factors gives
\begin{equation}
    \Delta = -\frac{U}{8N} \sum_k \frac{\Delta}{\sqrt{(\epsilon_{k+Q}+\epsilon_{-k+Q})^2/4+|\Delta|^2}} \left[ n(E^s_{k+Q,\uparrow}) +n(E^s_{-k+Q,\downarrow})-1 \right].
\end{equation}

\begin{figure}
    \centering
    \includegraphics{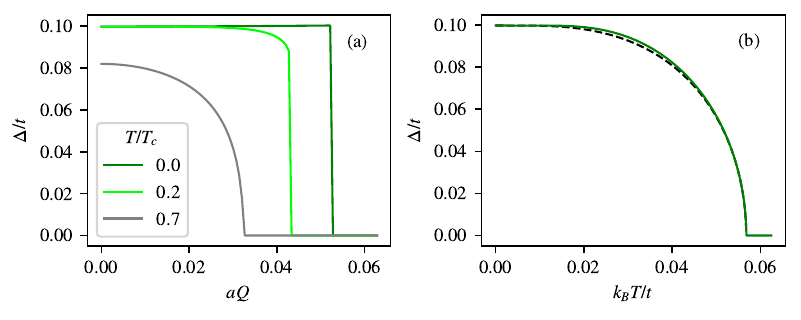}
    \caption{
    Solution to the gap equation for the spin singlet $s$-wave superconductor. In (a), the spin splitting is $h/t=0.01$, and in (b), $h=0$. The dashed black line in (b) is the approximate solution $\Delta^\sigma = \Delta_0 \tanh \left(1.74\sqrt{T_c/T-1  }\right)$ \cite{sigrist2005introduction}. The chemical potential is $\mu/t=-1$, and $U/t=5.32$. 
    }
    \label{fig:gapeq singlet}
\end{figure}
The gap equation is solved numerically in Fig. \ref{fig:gapeq singlet}. The supercurrent magnitudes used in Fig. 3(d) in the main text are normalized on the critical current $Q_c$ at which the superconducting order parameter drops to zero, see Fig. \ref{fig:gapeq singlet}(a).  The temperatures are normalized on the critical temperature $T_c\approx 0.057t/k_B$ at $Q=0$ and $h=0$, as shown in Fig. \ref{fig:gapeq singlet}(b).

\section{Tuning the supercurrents}
The momentum $Q$ of the Cooper pairs is related to the supercurrent $j$ through $j=-2ten_ea^2Q/\hbar$ for $|Qa|\ll1$ and at low temperature \cite{Takashima2017:PRB}. Here, $n_e$ is the density of superconducting electrons.
While there are several setups that will allow for charge and spin current injection into a superconductor, which are both converted into a supercurrent when triplet pairs are present, we provide one such example in Fig. \ref{fig:control supercurrents}. In (a), the impurity spins are placed on the surface of the superconductor and their ground state configuration is controlled by the supercurrent. In (b), an antiferromagnetic or altermagnetic insulating film is grown on top of the superconductor and the supercurrent controls the magnon gap. The superconductor is connected to two normal electrodes (gray), and the voltage $V_1$ controls the magnitude of the injected charge current $j_c$. This resistive current is converted to a supercurrent by the order parameter over the charge relaxation length $\lambda_Q$. The spin supercurrent is injected from the Hall bar (blue) by injection of a spin accumulation into the superconductor where the triplet order parameter converts it to a spin supercurrent. The voltage $V_2$ controls the magnitude of the transverse charge current and thus the spin supercurrent magnitude. As long as the contact between the Hall bar and the superconductor is smaller than $\lambda_Q$, the current will not be converted to a supercurrent in the superconductor and the Hall bar will therefore not be short-circuited. Moreover, the charge supercurrent $j_c$ will not escape into the Hall bar. If $V_1\neq 0$ and $V_2=0$, a pure charge supercurrent flows in the superconductor. If $V_1=0$ and $V_2\neq 0$, a pure spin supercurrent flows in the superconductor.

\begin{figure}
    \centering
    \includegraphics[width=0.75\linewidth]{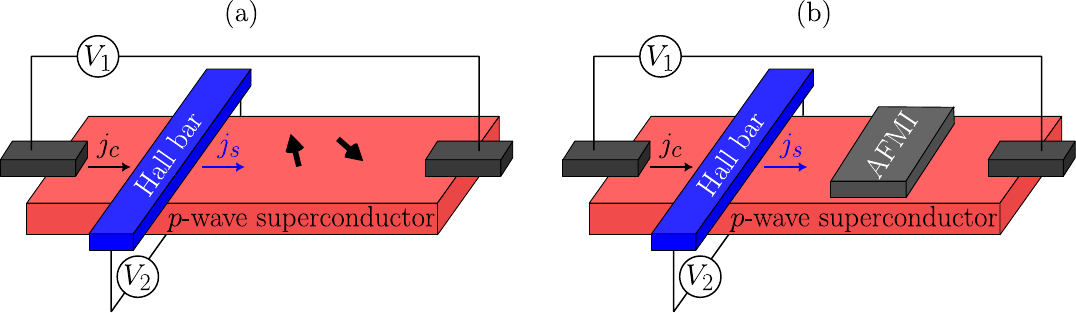}
    \caption{Proposed setup for controlling the supercurrent in the superconductor. In (a), the ground state configuration of two magnetic adatoms are controlled by the supercurrent. In (b), the magnon gap in a thin-film antiferromagnetic or altermagnetic insulator (AFMI) is controlled by the supercurrent. }
    \label{fig:control supercurrents}
\end{figure}

\end{document}